\documentclass[journal]{IEEEtran}
\IEEEoverridecommandlockouts

\usepackage[ruled,linesnumbered]{algorithm2e}

\usepackage[english]{babel}
\usepackage{blindtext}
    
\usepackage{CJKutf8}  

\usepackage{threeparttable}
\usepackage{pifont} % used for \ding{xx}

\newcommand{\cmark}{\textcolor{green!80!black}{\ding{51}}}
\newcommand{\xmark}{\textcolor{red}{\ding{55}}}

\usepackage{xcolor}
\usepackage{pgfplots}
\usepackage{tikz}
\usepackage{comment}
\usepackage{fontawesome}

\usepackage{filecontents}

\usepackage{url}
\usepackage{hyperref}

\usepackage{multirow}
\usepackage{amsmath}

\usepackage{textcomp}
\definecolor{ao(english)}{rgb}{0.0, 0.5, 0.0}

\usepackage{subfigure}
\usepackage{filecontents}
\usepackage{tikz}
\usetikzlibrary{positioning}

\usepackage{graphicx,subfigure}

\definecolor{bblue}{HTML}{4F81BD}
\definecolor{rred}{HTML}{C0504D}
\definecolor{ggreen}{HTML}{9BBB59}
\definecolor{pviolet}{HTML}{9F4C7C}

\usepackage{xspace}
\usepackage{listings}

\usepackage{color}  % Include the color package
\usepackage{amssymb}

\usepackage{etoolbox,xstring,mfirstuc,textcase}
\usepackage{booktabs}
\usepackage{pgfplots}
\newcommand{\inc}[1]{\textcolor{green!50!black}{↑ #1}}
\newcommand{\dec}[1]{\textcolor{red}{↓ #1}}

\usepackage{graphicx}   % for resizebox
\usepackage[table]{xcolor}   % for cellcolor in tables
\usepackage{makecell}   % optional, for multi-line cells
\usepackage{booktabs}   % optional, for nicer rules
\usepackage{array}   % for p{..} column types
\usepackage{tabularx}   % for automatic column width

\usepackage[most]{tcolorbox}
\newtcolorbox{insightbox}{
  enhanced,                              % REQUIRED for 'borderline'
  colback=gray!8,                       % light gray background
  frame hidden,                          % hide default frame
  borderline west={3pt}{0pt}{yellow!40!black}, % left blue stripe
  sharp corners,
  boxrule=0pt,                           % no regular border
  left=8pt, right=8pt, top=6pt, bottom=6pt,
  before skip=10pt, after skip=10pt
}

\usepackage{tabularx,booktabs,makecell,multirow, caption}
\usepackage{enumitem} % to control Itemization spacing

\newcolumntype{L}{>{\arraybackslash}X}

\usepackage{tikz}

\definecolor{findOptimalPartition}{HTML}{D7191C}
\definecolor{storeClusterComponent}{HTML}{FDAE61}
\definecolor{dbscan}{HTML}{ABDDA4}
\definecolor{constructCluster}{HTML}{2B83BA}

\begin{document}
%=================================================
\title{Does GenAI Rewrite How We Write? \\ An Empirical Study on Two-Million Preprints}
%=================================================   

\author{Minfeng Qi$^{1}$, Zhongmin~Cao$^{1}$, Qin Wang$^{2}$, Ningran Li$^{3}$, Tianqing Zhu$^{1\star}$\thanks{$^\star$Corresponding author. Email: \textit{tqzhu@cityu.edu.mo}. \\
}
\\ 
\smallskip
\textit{$^1$City University of Macau} $|$ \textit{$^2$CSIRO Data61} $|$ \textit{$^3$The University of Adelaide}
}

\begin{comment}

\author{Minfeng~Qi \textit{IEEE Member},
        Zhongmin~Cao, 
        Qin~Wang \textit{IEEE Member},
        Tianqing Zhu \textit{IEEE Senior Member}
        % Shiping Chen \textit{IEEE Senior Member}~
        % ~\IEEEmembership{Life~Fellow,~IEEE}% <-this % stops a space
\IEEEcompsocitemizethanks{\IEEEcompsocthanksitem M. Qi, Z. Cao, T. Zhu are affiliated with City University of Macau, MacauSAR, China. \protect
E-mail: \{mfqi, tqzhu\}@cityu.edu.mo

Q. Wang is affiliated with CSIRO Data61, Sydney, Australia. \protect
E-mail: \{Qin.Wang\}@data61.csiro.au

}
}
\end{comment}

%=================================================    
\maketitle
%=================================================  
\begin{abstract}
Preprint repositories become central infrastructures for scholarly communication. Their expansion transforms how research is circulated and evaluated before journal publication. Generative large language models (LLMs) introduces a further potential disruption by altering how manuscripts are written. While speculation abounds, systematic evidence of whether and how LLMs  reshape scientific publishing remains limited.
    
This paper addresses the gap through a large-scale analysis of more than 2.1 million preprints spanning 2016–2025 (115 months) across four major repositories (i.e., \textit{arXiv}, \textit{bioRxiv}, \textit{medRxiv}, \textit{SocArXiv}). We introduce a multi-level analytical framework that integrates interrupted time-series models, collaboration and productivity metrics, linguistic profiling, and topic modeling to assess changes in volume, authorship, style, and disciplinary orientation. Our findings reveal that LLMs have accelerated submission and revision cycles, modestly increased linguistic complexity, and disproportionately expanded AI-related topics, while computationally intensive fields benefit more than others. These results show that LLMs act less as universal disruptors than as selective catalysts, amplifying existing strengths and widening disciplinary divides. By documenting these dynamics, the paper provides the first empirical foundation for evaluating generative AI’s influence on academic publishing and highlights the need for governance frameworks that preserve trust, fairness, and accountability in an AI-enabled research ecosystem.

\end{abstract}

\begin{IEEEkeywords}
Generative AI, Large Language Models, Scholarly Communication, Preprints, Academic Publishing
\end{IEEEkeywords}

%=================================================
%\maketitle
%=================================================  

%-----------------------------------------
\section{Introduction}
%-----------------------------------------

The landscape of academic publishing has undergone rapid structural changes over the past decade. On one front, preprint repositories (e.g., arXiv and bioRxiv) that operate with fewer gatekeeping constraints have become a central component of scholarly communication, allowing researchers to share manuscripts prior to peer review. These repositories provide an open-access stream of academic output, reflecting evolving practices in writing and dissemination at scale.

In parallel, generative AI (GenAI)~\cite{bommasani2021opportunities}, driven by advances in transformer-based architectures~\cite{achiam2023gpt,openai2024gpt4o}, has made large language models (LLMs) capable of producing coherent academic prose with minimal prompting. The release of GPT-3 initiated the first wave of AI-assisted writing, which rapidly expanded with ChatGPT and other subsequent models. 

The growing integration of GenAI into research practice raises foundational questions about how scientific knowledge is produced and communicated. AI systems can accelerate writing and revision, alter collaboration structures, and potentially reshape linguistic and disciplinary conventions. Yet, despite widespread normative responses, ranging from publisher disclosure policies~\cite{science2023ai,elsevier2023ai,taylor2023ai} to institutional guidelines at leading universities~\cite{harvard2023guidelines,columbia2023policy,unc2023guidance}, systematic empirical evidence on how GenAI is influencing scholarly production remains limited. This motivates the \textbf{central question} of our study: \textit{Has GenAI merely accelerated scientific writing—or has it begun to reorganize how knowledge itself is produced?}

Rather than treating these effects as isolated outcomes, we approach them as an interconnected sequence of transformations: from the acceleration of research cycles to the reorganization of scholarly labor, the redirection of research focus, and the eventual recomposition of disciplinary boundaries.

{\setlength{\leftmargini}{3em}
\begin{itemize}
    \item[\textcolor{magenta}{RQ1:}] \textit{Acceleration.} Has generative AI accelerated the pace of scientific publishing?
    \item[\textcolor{magenta}{RQ2:}] \textit{Reorganization.} Has it reshaped how researchers collaborate and produce manuscripts?
    \item[\textcolor{magenta}{RQ3:}] \textit{Expression.} Has it changed the language and style of academic writing?
    \item[\textcolor{magenta}{RQ4:}] \textit{Redirection.} Has it redirected research attention toward AI-related themes?
    \item[\textcolor{magenta}{RQ5:}] \textit{Recomposition.} Has it altered the disciplinary balance of scientific output?
\end{itemize}}

% To operationalize this inquiry, we examine five interrelated dimensions of change:

% \begin{itemize}
%     \item Submission volumes and growth rates.
%     \item Collaboration patterns and authorship structures.
%     \item Linguistic style and readability shifts.
%     \item Topical emergence and diffusion.
%     \item Disciplinary reconfiguration.
% \end{itemize}

% These dimensions jointly examine the quantitative, linguistic, and structural dimensions of GenAI’s impact on the scientific ecosystem. They aim to capture whether large-scale shifts in academic behavior accompany the adoption of generative tools—and if so, whether such shifts are uniform across repositories and disciplines.

A limited number of studies have begun probing this intersection (\textcolor{teal}{Table~\ref{tab:comparative}}). Survey-based and bibliometric works~\cite{kalla2023study,zamfiroiu2023chatgpt} primarily catalog publication trends and reference patterns, whereas commentaries~\cite{haman2024using} rely on anecdotal reflections about GenAI in scholarly workflows. Parallel works explore automation and agent-based research~\cite{ifargan2025autonomous,schmidgall2025agent}, as well as LLM evaluation in peer review~\cite{zhou2024llm} and stylistic bias analysis~\cite{malik2024empirical}. However, these studies remain narrow in scope, limited in sample size, or speculative in focus. To the best of our knowledge, no prior work has conducted a large-scale, cross-disciplinary, and data-driven analysis of how GenAI reshapes the global dynamics of scientific publishing.

\begin{table}[!hbt]
\vspace{-0.05in}
\centering
\caption{This Work \textcolor{gray}{v.s.} Prior GPT\&Research Studies}\label{tab-gpt-compare}
\label{tab:comparative}
\vspace{-0.03in}
\resizebox{\linewidth}{!}{
\begin{tabular}{c|cc|cc}
\toprule
\textit{\textbf{Exa.}} & \textit{\textbf{Method}} & \textit{\textbf{Target}} & \textit{\textbf{Scope}} & \textit{\textbf{Scale}} \\
\midrule
\midrule
\cite{kalla2023study} & Literature review & Fields of study & \textcolor{red}{Generic} & \textcolor{red}{$<$50} \\

\cite{zamfiroiu2023chatgpt} & Systematic review & Published papers  & \textcolor{red}{Generic} & \textcolor{red}{$<$300} \\

\cite{haman2024using} & Commentary & Literature workflow & \textcolor{red}{Narrow} & \textcolor{red}{$<$5} \\

\cite{ifargan2025autonomous} & Agent simulation & Auto-research agents  & \textcolor{red}{Narrow} & \textcolor{red}{Synthetic} \\

\cite{schmidgall2025agent} & Framework + Sim & Agent behavior & \textcolor{red}{Niche} & \textcolor{red}{$<$30} \\

\cite{zhou2024llm} & Benchmark eval & Peer review task & \textcolor{red}{Single task} & \textcolor{red}{Small-scale} \\

\cite{malik2024empirical} & Stylometric analysis & Writing style  & \textcolor{red}{Bias-focused} & \textcolor{red}{$<100$ outputs} \\

\cmidrule{1-5}

\textit{\textbf{Ours}} & Empirical study & Preprint corpus & \textcolor{blue}{Cross-disciplinary} & \textcolor{blue}{$>$2M papers} \\

\bottomrule
\end{tabular}
}
\vspace{-0.05in}
\end{table}

\smallskip
\noindent\textbf{Our approach.}
Our analysis draws on a large-scale dataset compiled from four major preprint repositories: arXiv, bioRxiv, medRxiv, and SocArXiv. The collection spans the period from January 2016 through August 2025 (lasting 115 months) and contains more than 2.1 million manuscripts. 

\begin{figure}[!t]
  \centering
  \includegraphics[width=0.95\linewidth]{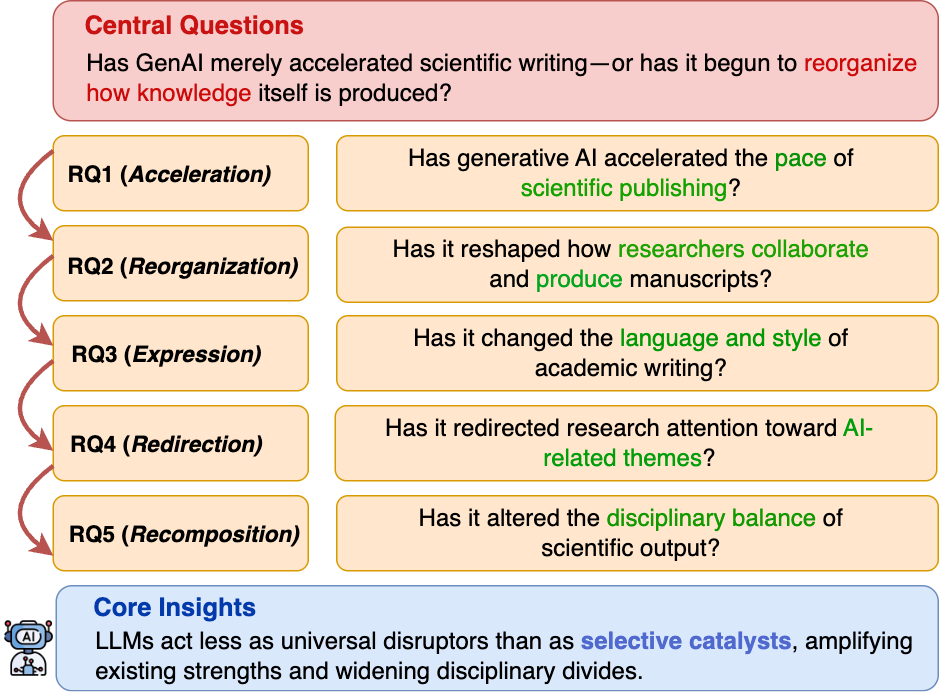}
  \caption{Overview of research questions}
  \label{fig:RQs}
\end{figure}

Building on this corpus, we address the central research question through a layered logic that moves from system-wide signals to field-specific nuance (as detailed in \textcolor{teal}{Fig.\ref{fig:RQs}}). 

First, we investigate whether the overall scale of scholarly production has shifted in the wake of major LLM milestones (e.g., GPT-3 and ChatGPT) by analyzing submission volumes and growth trajectories across repositories (\textcolor{magenta}{RQ1}). Building on these system-level trends, we then examine how patterns of collaboration and authorship have evolved, focusing on changes in team size, author productivity, and revision tempo (\textcolor{magenta}{RQ2}). Moving from research organization to linguistic expression, we assess whether the textual characteristics of academic manuscripts have been affected by GenAI, through measures of readability and linguistic complexity. (\textcolor{magenta}{RQ3}). We further explore whether these transformations are accompanied by a thematic reorientation of research agendas, tracing the emergence and diffusion of AI-related topics identified through keyword analysis and topic modeling (\textcolor{magenta}{RQ4}). Finally, we evaluate how these effects vary across disciplines, comparing the pace and magnitude of adoption to reveal where the influence of GenAI has been most pronounced (\textcolor{magenta}{RQ5}).

% First, we assess whether submission volumes and growth rates exhibit structural breaks following the release of GPT-3 in 2020 and ChatGPT in 2022 (\textcolor{magenta}{RQ1}). Second, we examine collaboration and productivity patterns, analyzing average team size, the proportion of single-author papers, author-level publication rates, and revision intervals (\textcolor{magenta}{RQ2}). Third, we evaluate linguistic features of scientific text, focusing on readability scores, syntactic complexity, and vocabulary usage (\textcolor{magenta}{RQ3}). Fourth, we investigate thematic dynamics through topic modeling and keyword analysis, tracing the emergence and growth of AI-related themes across repositories (\textcolor{magenta}{RQ4}). Finally, we analyze disciplinary heterogeneity by comparing adoption trajectories across domains, identifying which fields exhibit the strongest signals of change (\textcolor{magenta}{RQ5}). 

\smallskip
\noindent\textbf{Contributions.}  
We present the first large-scale empirical study of how generative language models influence academic publishing. We collect and analyse over two million preprints across four major repositories and uncover a series of measurable signals of transformation in academic publishing. 
The key contributions are as follows:

\noindent\hangindent 1em\textit{$\triangleright$ A multi-level analytical framework for measuring AI's impact on scholarly publishing} (\textcolor{teal}{\$\ref{sec:framework}}). We propose a structured methodology that integrates interrupted time-series models, productivity metrics, linguistic profiling, and topic modeling. This framework enables a systematic investigation of the volume, style, and thematic evolution of research outputs in response to generative AI tools. It is applicable to other scientific corpora beyond preprints and provides a foundation for AI-mediated knowledge production.

\noindent\hangindent 1em\textit{$\triangleright$ A large-scale, open dataset of preprints annotated with rich metadata and linguistic features} (\textcolor{teal}{\$\ref{sec:data}}).
We construct and release a curated dataset comprising over 2.1 million manuscripts from arXiv, bioRxiv, medRxiv, and SocArXiv, covering the period from 2016 to 2025. The dataset includes derived features for author metadata, readability scores, lexical richness, etc., facilitating reproducible research on scientific communication at scale.

\noindent\hangindent 1em\textit{$\triangleright$ Empirical findings on the structural and stylistic transformations in research publishing post-ChatGPT} (\textcolor{teal}{\$\ref{sec:RQ1}-\ref{sec:RQ5}}).  
Our analyses reveal statistical shifts in submission and revision tempos (RQ1–RQ2), measurable increases in linguistic complexity in abstracts (RQ3), disproportionate expansion of AI-related research topics across repositories (RQ4), and selective acceleration in computationally intensive disciplines relative to slower-adopting fields (RQ5). 

We conclude that the influence of GenAI on academic publishing is real but uneven, acting as a catalyst in some areas while leaving others largely unaffected.

\noindent\hangindent 1em\textit{$\triangleright$ A structured set of practical implications for governance} (\textcolor{teal}{\$\ref{sec:implications}}). We further develop a forward-looking framework that translates observed patterns into practical guidance.

%============================================
\section{Background and Foundations}
\label{sec:background}
%============================================

\subsection{Evolution of LLMs}
LLMs has progressed rapidly over the past five years. OpenAI's GPT (Generative Pre-trained Transformer) series has played a central role in this evolution. GPT-2, released in 2019, demonstrated that a large-scale transformer model trained on a diverse corpus could generate surprisingly coherent text~\cite{radford2019language}. GPT-3, introduced in 2020, dramatically scaled up the parameter count to 175 billion and showcased few-shot and zero-shot capabilities across a range of tasks~\cite{brown2020language}.

However, GPT-3 remained largely accessible only via API and lacked alignment with human intent. This limitation was addressed by InstructGPT~\cite{ouyang2022training}, a fine-tuned variant trained using reinforcement learning from human feedback, which improved the model’s ability to follow user instructions. Building on this, OpenAI released ChatGPT in November 2022, based on GPT-3.5, which introduced a conversational interface that made LLMs widely usable by non-experts. Its successor, GPT-4, launched in March 2023, offered improved reasoning ability and multimodal support~\cite{openai2023gpt4}.

Concurrently, other organizations developed competitive LLMs such as Anthropic’s Claude~\cite{claude2023}, Google DeepMind’s Gemini (formerly Bard)~\cite{google2023gemini}, Meta’s LLaMA~\cite{touvron2023llama}, and DeepSeek~\cite{liu2024deepseek}. These models vary in training data scale, instruction tuning methodology, and openness, but collectively reflect a global trend toward the rapid advancement foundation models. The recent release of GPT-4o in May 2024~\cite{openai2024gpt4o} and GPT-5~\cite{openai2025gpt5}, which supports seamless integration of text, audio, and vision inputs, marks the emergence of fully multimodal assistants with near-real-time interaction capabilities.

\begin{figure}[!h]
  \centering
  \includegraphics[width=\linewidth]{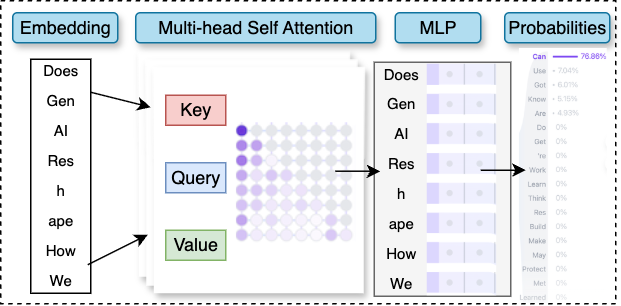}
  \caption{LLM transformer architecture}
  \label{fig:transformer}
\end{figure}

\subsection{Technical Foundations of LLMs}
Most current LLMs are built upon the Transformer architecture (\textcolor{teal}{Fig.\ref{fig:transformer}}), a deep learning model introduced by Vaswani et al.~\cite{vaswani2017attention}. The Transformer fundamentally changed the field of natural language processing by introducing the self-attention mechanism, enabling models to capture long-range dependencies in sequences without relying on recurrence. This architecture has since become the foundation for most state-of-the-art generative models across text, code, image, and multimodal domains~\cite{wolf2020transformers,qi2025towards}.

At its core, the Transformer operates by predicting the next token in a sequence based on prior context, using stacked layers of multi-head self-attention and feed-forward networks. The model ingests input text through an embedding layer that converts tokens into high-dimensional vectors. These embeddings are enriched with positional encodings, allowing the model to infer the order of tokens.

Each Transformer block contains two primary components: (1) a \textit{multi-head self-attention} mechanism that enables the model to focus on different parts of the sequence simultaneously, and (2) a \textit{multi-layer perceptron} (MLP) that refines the token representations. Residual connections and layer normalization are applied throughout to improve gradient flow and training stability~\cite{han2021transformer}. The output of the final block is then passed through a softmax layer over the vocabulary to generate probability distributions over potential next tokens.

\subsection{Prompt Engineering and Instruction Tuning}
\label{sec:prompt-instruct}
Prompt engineering refers to the practice of designing input sequences that condition the model to perform a desired task without modifying its internal parameters. Early studies demonstrated that LLMs exhibit emergent few-shot capabilities, wherein performance improves when provided with a handful of examples formatted in a consistent pattern~\cite{brown2020language}. This led to the widespread adoption of zero-shot, few-shot, and chain-of-thought prompting techniques~\cite{allingham2023simple,yuan2024instance,abdul2023align,hossain2024visual}. The latter, in particular, encourages models to verbalize intermediate reasoning steps and has been shown to enhance accuracy on complex reasoning tasks~\cite{li2025structured,shao2023synthetic}. More advanced prompting strategies include self-consistency decoding~\cite{chen2024self} and least-to-most prompting~\cite{ma2025should}.

In parallel, instruction tuning has provided a scalable mechanism for transforming generic LLMs into instruction-following models capable of adhering to user goals with higher precision. This process typically involves supervised fine-tuning on a curated set of instruction-response pairs, followed by RLHF. The InstructGPT framework~\cite{ouyang2022training} formalized this pipeline, showing that even a smaller fine-tuned model could outperform larger base models on alignment-sensitive tasks. Subsequent efforts, such as SFT~\cite{yang2024s}, Lofit~\cite{yin2024lofit}, and self-play fine turning~\cite{yuan2024self}, have emphasized the diversity of datasets to reduce dependency on proprietary data.

%============================================
\section{Methodology}
\label{sec:methodology}
\begin{figure}[!t]
  \centering
  \includegraphics[width=\linewidth]{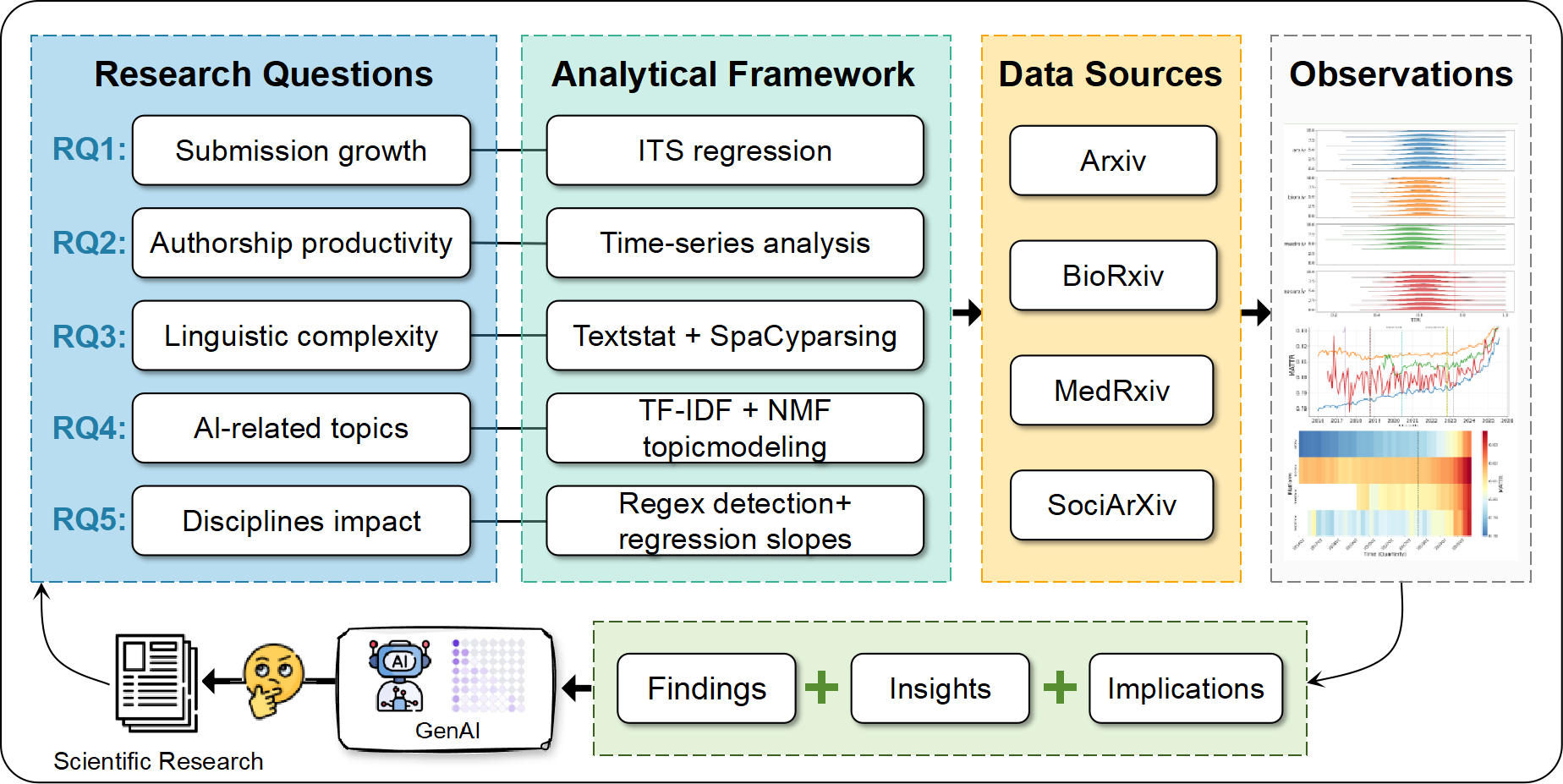}
  \caption{Methodology}
  \label{fig:methodology}
\end{figure}

%============================================
We provide an overview of the methodological framework (\textcolor{teal}{Fig.\ref{fig:methodology}}), outlining the data collection and analytical techniques.

\subsection{Corpus Construction}
\label{sec:data}

\begin{figure}[!t]
  \centering
  \includegraphics[width=0.85\linewidth]{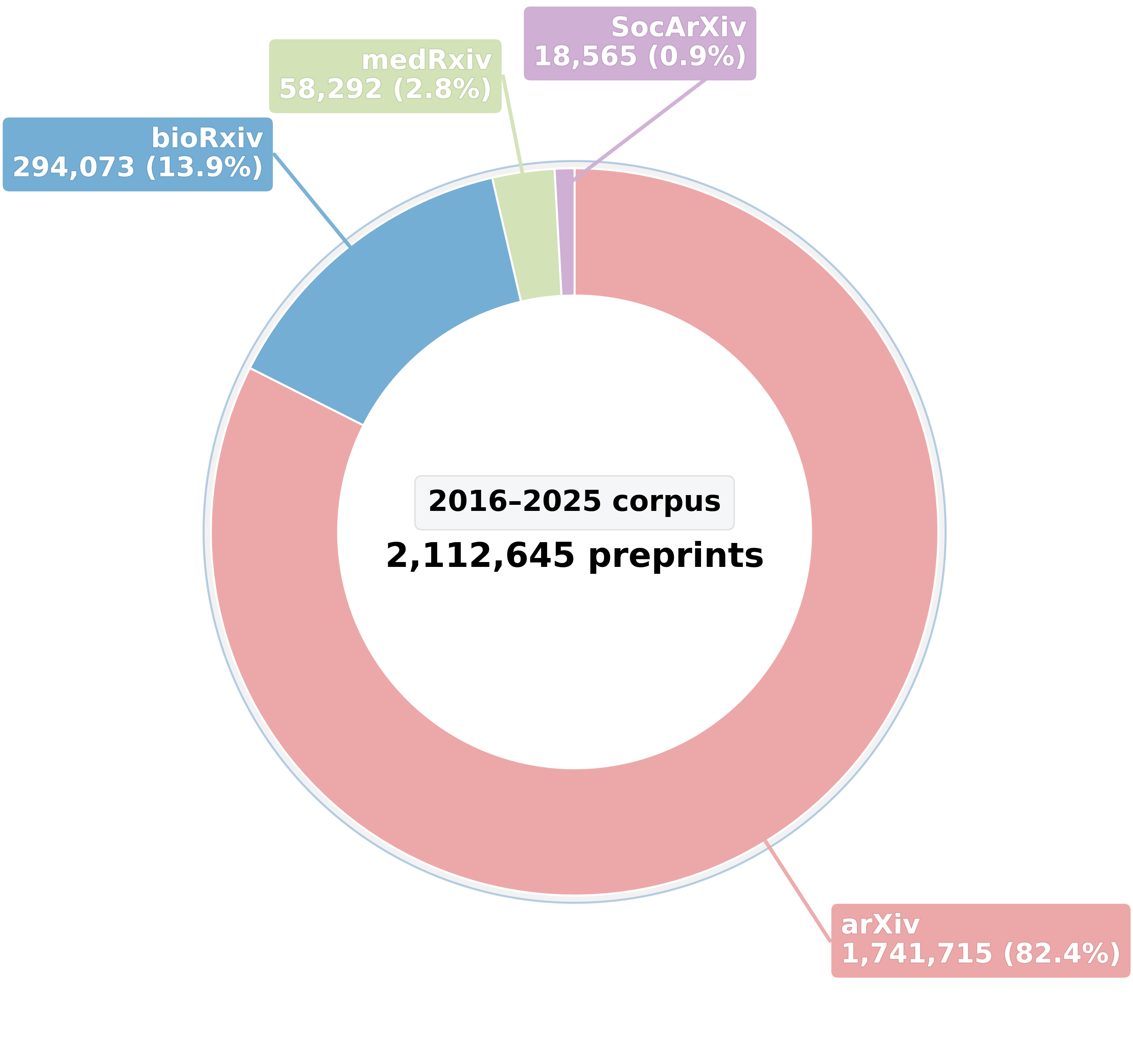}
  \caption{Composition of the 2016–2025 corpus across repositories. The dataset integrates four major preprint platforms: \textit{arXiv}, \textit{bioRxiv}, \textit{medRxiv}, and \textit{SocArXiv}.}
  \label{fig:dataset_composition}
\end{figure}

\noindent\textbf{Data collection.}  
Our analyses draw on a multi-repository corpus of preprints spanning the period from 2016 to 2025, covering the natural sciences, life sciences, medicine, and social sciences (\textcolor{teal}{Fig.\ref{fig:dataset_composition}}). The dataset integrates four major platforms: \textit{arXiv}, \textit{bioRxiv}, \textit{medRxiv}, and \textit{SocArXiv}. Each repository was accessed through its official programmatic interface\footnote{arXiv API: \url{https://export.arxiv.org/oai2};  
bioRxiv API: \url{https://api.biorxiv.org/details/biorxiv};  
medRxiv API: \url{https://api.biorxiv.org/details/medrxiv};  
socArxiv API: \url{https://api.osf.io/v2/preprints}.}  
to ensure reproducibility and avoid reliance on third-party scrapers that may introduce inconsistencies.

In constructing the corpus, we focused on metadata fields that are both universally available and analytically relevant. Specifically, for each manuscript we extracted identifiers, titles, abstracts, subject categories, submission and update dates, and full author lists. These fields jointly enable the operationalization of submission dynamics (\textcolor{magenta}{RQ1}), collaboration patterns (\textcolor{magenta}{RQ2}), linguistic features (\textcolor{magenta}{RQ3}), topical distributions (\textcolor{magenta}{RQ4}), and disciplinary variation (\textcolor{magenta}{RQ5}). By restricting the analysis to abstracts rather than full-text content, we ensured cross-platform comparability while minimizing the risks of inconsistent text coverage and repository-specific biases.

\vspace{-0.1in}
\begin{center}
\resizebox{\linewidth}{!}{
\fbox{%
\begin{minipage}{0.999\linewidth}

          %\faWarning~
          \textbf{\text{Ethical considerations:}}  All data were collected from openly accessible preprint repositories via their official APIs, in full compliance with the platforms' terms of service. No personally identifiable information beyond publicly listed author names was retrieved, and all analyses were conducted at the aggregate level. Potential biases from incomplete metadata, heterogeneous platform coverage, or uneven disclosure practices are acknowledged and discussed as methodological limitations.

\end{minipage}
}
}
\end{center}

\smallskip
\noindent\textbf{Data processing.} All raw records were processed through a unified cleaning pipeline designed to maximize internal consistency across repositories. Steps included: (i) standardization of date formats into ISO 8601 for temporal comparability; (ii) normalization of subject labels through a controlled vocabulary to mitigate taxonomic inconsistencies; (iii) disambiguation of author names using heuristic rules on name ordering, punctuation, and affiliation markers; and (iv) removal of invalid or incomplete entries. Deduplication procedures were applied to handle cross-posted manuscripts, ensuring that multiple deposits of the same work were collapsed into a single record. Finally, for repositories with occasional missing metadata fields, we applied controlled imputation strategies based on empirical cross-platform distributions, thereby preserving realistic variance while avoiding artificial homogenization.

\subsection{Analytical Framework}
\label{sec:framework}
To answer the above research questions, we designed a set of complementary analyses tailored to each dimension of interest. Each RQ is discovered with specific data sources, metrics, and statistical approaches, as detailed in \textcolor{teal}{Table.\ref{tab:result-guidance}}.

\begin{table}[!htb]
\centering
\caption{(Result guidance) Mapping RQs to Figures}
\label{tab:result-guidance}
\scriptsize
\renewcommand{\arraystretch}{1.1}
\begin{tabular}{c|c|l@{}}
\toprule
\multicolumn{1}{c}{\textbf{RQ}} & \multicolumn{1}{c}{\textbf{Index}} & \multicolumn{1}{c}{\textbf{Brief description}} \\
\midrule
\midrule
\multirow{4}{*}{\textbf{\textcolor{magenta}{RQ1}}} 
& \textcolor{teal}{Fig.\ref{fig:rq1_monthly}}    & Monthly submissions with MA(3) smoothing \\
& \textcolor{teal}{Fig.\ref{fig:rq1_cumulative}} & Cumulative submission growth across repositories \\
& \textcolor{teal}{Fig.\ref{fig:rq1_yoy}}        & Year-over-year growth rate comparison \\
& \textcolor{teal}{Table~\ref{tab:rq1-its}}       & ITS regression estimates for LLM milestones \\
\midrule

\multirow{4}{*}{\textbf{\textcolor{magenta}{RQ2}}} 
& \textcolor{teal}{Fig.\ref{fig:rq2_mean_authors}}  & Average team size by repository \\
& \textcolor{teal}{Fig.\ref{fig:rq2_step2}}         & Author productivity (papers per year) \\
& \textcolor{teal}{Fig.\ref{fig:rq2_step3_density}} & Distribution of revision intervals \\
& \textcolor{teal}{Fig.\ref{fig:rq2_step3_line}}    & Annual revision tempo with LLM markers \\
\midrule

\multirow{6}{*}{\textbf{\textcolor{magenta}{RQ3}}} 
& \textcolor{teal}{Fig.\ref{fig:fkgl_4sites}}   & Flesch–Kincaid grade levels over time \\
& \textcolor{teal}{Fig.\ref{fig:fre_4sites}}    & Flesch Reading Ease scores over time \\
& \textcolor{teal}{Fig.\ref{fig:sentlen}}       & Average sentence length trends \\
& \textcolor{teal}{Fig.\ref{fig:difficult}}     & Difficult-word ratio across repositories \\
& \textcolor{teal}{Fig.\ref{fig:lexical}}       & Lexical density (content word share) \\
& \textcolor{teal}{Fig.\ref{fig:subordination}} & Subordination ratio (syntactic depth) \\
\midrule

\multirow{2}{*}{\textbf{\textcolor{magenta}{RQ4}}} 
& \textcolor{teal}{Fig.\ref{fig:rq5_aishare}}    & AI-related document share by year \\
& \textcolor{teal}{Fig.\ref{fig:rq5_wordclouds}} & Topic modeling word clouds by domain \\
\midrule

\multirow{6}{*}{\textbf{\textcolor{magenta}{RQ5}}} 
& \textcolor{teal}{Fig.\ref{fig:arxiv_ridge}}    & arXiv subject-level AI adoption \\
& \textcolor{teal}{Fig.\ref{fig:biorxiv_ridge}}  & bioRxiv subject-level AI adoption \\
& \textcolor{teal}{Fig.\ref{fig:medrxiv_ridge}}  & medRxiv subject-level AI adoption \\
& \textcolor{teal}{Fig.\ref{fig:socarxiv_ridge}} & SocArXiv subject-level AI adoption \\
& \textcolor{teal}{Fig.\ref{fig:combined_density}} & Cross-field deltas and post-ChatGPT slopes \\
& \textcolor{teal}{Fig.\ref{fig:combined_trends}}  & Field-specific adoption trajectories \\
\bottomrule
\end{tabular}
\end{table}

\begin{table*}[!htbp]
\centering
\caption{Quick access to our analytical framework}
\label{tab:rq_methods}
\resizebox{\linewidth}{!}{
\begin{tabular}{c|l|l|l}
\toprule
  \multicolumn{1}{c}{\textbf{RQ}} & \multicolumn{1}{c}{\textbf{Data}} & \multicolumn{1}{c}{\textbf{Metrics}} & \multicolumn{1}{c}{\textbf{Methods}} \\
\midrule
\midrule
\textcolor{magenta}{RQ1} & Monthly submissions & Counts, cumulative totals, growth rates & ITS regression, seasonal controls, COVID intervention \\
\textcolor{magenta}{RQ2} & Author metadata & Team size, solo share, per-author output, Gini/Lorenz, revision tempo & Time-series analysis, distributional inequality \\
\textcolor{magenta}{RQ3} & Title + abstract & FRE, FKGL, sentence length, lexical density, subordination ratio, difficult word ratio & textstat, spaCy parsing, monthly aggregation \\
% \textcolor{magenta}{RQ4} & Titles + abstracts & TTR, MATTR, JSD (n-grams), cosine similarity (TF–IDF) & Stylometric comparison, temporal divergence analysis \\
\textcolor{magenta}{RQ4} & Title + abstract + topic label & AI-topic share, keyword hits, topic distributions & TF–IDF + NMF topic modeling, keyword regex, OLS trend \\
% \textcolor{magenta}{RQ6} & Titles + abstracts & Method-class lexicon (200+ terms), quantitative/qualitative/mixed shares & Rule-based + topic-enhanced classification, monthly trends \\
\textcolor{magenta}{RQ5} & Title + abstract + subject  & AI keyword hits, adoption month, growth slopes, field-normalized share & Regex detection, threshold analysis, regression slopes, heatmaps \\
\bottomrule
\end{tabular}
}
\end{table*}

\smallskip
\noindent{\ding{228} \textbf{\textcolor{magenta}{RQ1}: Submission volumes and growth rates.}}  
A central concern is whether LLMs have altered the scholarly output. To probe this, we focus on submission activity through three complementary indicators: absolute monthly counts, cumulative totals, and annualized growth rates. While raw counts capture scale, growth rates normalize for repository size and highlight acceleration or deceleration trends. To formally test for structural breaks, we employ interrupted time-series (ITS) regression models~\cite{mcdowall2019interrupted} of the form  
\[
\begin{aligned}
Y_t = \;& \beta_0 + \beta_1 t 
    +\beta_2 D_{\text{GPT3}}+\beta_3 (t \times D_{\text{GPT3}}) \\
     &+\beta_4 D_{\text{ChatGPT}}+\beta_5 (t \times D_{\text{ChatGPT}}) \\
     &+\beta_6 D_{\text{COVID}}+\epsilon_t
\end{aligned}
\]

Here, \(Y_t\) represents the number of submissions observed at time \(t\), while \(t\) itself captures the underlying linear time trend. The intervention indicator \(D_{\text{GPT3}}\) is defined as a binary variable that equals 1 for months after June 2020 (the release of GPT-3) and 0 otherwise, and \(D_{\text{ChatGPT}}\) similarly marks the period after November 2022 (the release of ChatGPT). The interaction terms \(t \times D_{\text{GPT3}}\) and \(t \times D_{\text{ChatGPT}}\) allow the slope of the submission trajectory to change following these interventions. The variable \(D_{\text{COVID}}\) denotes the pandemic period (March 2020–June 2021), accounting for exogenous disruptions unrelated to LLM developments. Coefficients \(\beta_0\) and \(\beta_1\) capture the baseline level and secular growth rate of submissions, while \(\beta_2\) and \(\beta_4\) measure immediate level shifts at the GPT-3 and ChatGPT interventions, and \(\beta_3\) and \(\beta_5\) estimate changes in slope following these events. The parameter \(\beta_6\) isolates the effect of COVID-19 disruptions. Finally, \(\epsilon_t\) is the error term, modeled with HAC-robust standard errors to mitigate serial correlation. This design isolates whether observed inflections are attributable to LLM milestones rather than background temporal trends, providing a rigorous evaluation of ChatGPT’s impact on submission trajectories.

\smallskip
\noindent{\ding{228} \textbf{\textcolor{magenta}{RQ2}: Collaboration patterns and author productivity.}}
Patterns of authorship offer a window into how research is organized. We assess collaboration through two indicators: the \textit{mean number of authors per paper}, which captures the extent of teamwork, and the \textit{proportion of single-author papers}, which reflects the persistence of individual research. Productivity is assessed at the author level using the average number of publications per year, providing a measure of overall research output, and distributional inequality is quantified through the Gini coefficient and Lorenz curves~\cite{gastwirth1972estimation}. 

The Gini coefficient ranges from 0 (perfect equality) to 1 (maximum inequality), summarizing whether a small group of authors dominates output, while the Lorenz curve visualizes the cumulative distribution of publications across the author population. To further capture efficiency, we compute revision tempo, defined as the inter-submission interval (\( \text{interval\_days} = t_{i+1} - t_i \)) for consecutive submissions by the same author, with shorter gaps indicating accelerated cycles of research and dissemination.

\smallskip
\noindent{\ding{228} \textbf{\textcolor{magenta}{RQ3}: Readability and linguistic complexity.}}
This question evaluates whether the linguistic texture of academic manuscripts shifted, focusing on both readability and syntactic complexity. Readability is assessed using two standard indices: the Flesch Reading Ease (FRE)~\cite{farr1951simplification}, which ranges from 0 to 100 with higher scores indicating easier comprehension, and the Flesch-Kincaid Grade Level (FKGL)~\cite{solnyshkina2017evaluating}, which estimates the U.S. school grade required to understand a text. Linguistic complexity is captured through four complementary metrics: (i) average sentence length (in words), which reflects syntactic elaboration; (ii) lexical density, defined as the proportion of content words (nouns, verbs, adjectives, adverbs) over total words, capturing informational richness; (iii) subordination ratio, measuring the frequency of subordinate clause markers as an indicator of syntactic depth; (iv) and difficult word ratio, quantifying the share of tokens outside standard frequency lists, thus approximating vocabulary difficulty.  

To compute these indicators, titles and abstracts were processed using {\small\textsf{spaCy}} for sentence segmentation, tokenization, and dependency parsing, while readability indices were derived with {\small\textsf{textstat}}. Invalid or incomplete records were removed, dates were standardized, and all metrics were aggregated to the monthly level for comparability across repositories. Analytical procedures included time-series analysis to track temporal trajectories from 2016 to 2023, mean comparisons before and after the ChatGPT release, and correlation analysis to evaluate convergence across repositories.

\smallskip
\noindent{\ding{228} \textbf{\textcolor{magenta}{RQ4}: Emergence of AI-related topics.}} 
Topical content provides another dimension of change: are AI-related research areas expanding disproportionately in the LLM era? To address this, we rely on both keyword-based detection and topic modeling applied to titles and abstracts. At the document level, each record is assigned binary indicators for whether it is AI-related (\textit{is\_ai\_doc}) and whether it belongs to an AI-focused topic cluster (\textit{is\_ai\_topic}). The overall prevalence of such work is summarized by the monthly share of AI-related documents (\textit{ai\_share}), defined as the ratio of AI-flagged papers to the total number of submissions.  

To capture latent thematic structures, we apply a TF–IDF representation~\cite{zhang2011comparative} with a vocabulary cap of 200,000 features and perform Non-negative Matrix Factorization (NMF)~\cite{lee2000algorithms} to extract 120 topics. Each topic is characterized by its identifier (\textit{topic\_id}), associated top keywords, and document membership scores. Topics are labeled by inspecting their most frequent terms, and AI-related themes are detected through a dual mechanism: keyword regular expressions (e.g., “ChatGPT,” “LLM,” “transformer”) and manual validation. This minimises false positives and ensures that both explicit mentions and latent thematic clusters are captured.  

Temporal dynamics are modelled at the monthly level (\textit{ym}), with additional smoothing via three-month moving averages to reduce volatility. We evaluate whether AI-related topics exhibit significant growth post-ChatGPT using statistical tests, including independent-sample \textit{t}-tests to compare mean adoption rates before and after November 2022, and ordinary least squares (OLS) regressions to estimate trend slopes.

\smallskip
\noindent\textbf{\ding{228} \textcolor{magenta}{RQ5}: Disciplinary differences.}  
The impact of LLMs can hardly be uniform across domains. We stratify the corpus using repository-specific subject taxonomies, including arXiv domain labels, bioRxiv and medRxiv classifications, and OSF categories for SocArXiv. Each is tagged with \textit{primary\_subject}, enabling aggregation of indicators at the field level.  

The central outcome variable is the share of AI-related manuscripts (\textit{share}), defined as the proportion of documents per discipline containing AI-related keywords in titles or abstracts. To capture temporal dynamics, we compute both monthly (\textit{ym}) and annual (\textit{year}) aggregates, with smoothing via three-month moving averages (\textit{share\_ma3}). Pre-/post-ChatGPT contrasts are quantified using average adoption rates (\textit{pre\_mean}, \textit{post\_mean}) and their difference (\textit{delta}), while linear trends are estimated through post-release regression slopes (\textit{slope\_post}). Adoption thresholds are further identified by the first month in which a discipline surpasses a 1\% AI-share benchmark (\textit{adoption\_month}).  

% These measures allow us to disentangle baseline disciplinary differences from ChatGPT-specific effects. For example, a large \(\delta\) combined with an early \textit{adoption\_month} signals accelerated adoption in a given field, while flat slopes may indicate limited responsiveness. Comparative analyses employ field-normalized indices to mitigate disparities in repository size, ensuring that growth in smaller disciplines is not masked by scale effects in larger ones. 

%============================================
\section{Experiments}
\label{sec:expriments}
%============================================

\subsection{Experimental Setup} 
\label{sec:setup}
\noindent\textbf{Data scale.}
We conducted all analyses on a large-scale corpus of approximately \underline{2.11 million} preprints spanning the period from January 2016 through August 2025. 

The dataset integrates four major repositories that cover the natural sciences, life sciences, medicine, and social sciences: \textit{arXiv}, \textit{bioRxiv}, \textit{medRxiv}, and \textit{SocArXiv}. Within this 2016–2025 window, the counts are: \textit{arXiv} 1{,}741{,}715 new submissions (arXiv’s cumulative size grew from 576{,}864 in Jan 2016 to 2{,}318{,}579 by Aug 2025), \textit{bioRxiv} 294{,}073, \textit{medRxiv} 58{,}292 (since 2019), and \textit{SocArXiv} 18{,}565 (since 2018), totaling 2{,}112{,}645 manuscripts used for analysis.

% \qw{comments} %这里data计算的总数不对吧 2.11m total 但是arxiv一个就2.3m？ 如果比如说是因为arxiv存在时间早于datacapture时间 要加个注释说一下时间线

For consistency across platforms, we restricted our analyses to metadata fields that were universally available: \textit{titles}, \textit{abstracts}, \textit{author lists}, \textit{submission dates}, and \textit{subject classifications}. After collection, all records underwent a standardized preprocessing pipeline that included normalization of date formats, parsing and disambiguation of author names, deduplication of overlapping entries, and exclusion of incomplete or corrupted records. Implausible values, such as negative revision intervals, were removed, and occasional missing metadata was imputed using controlled cross-platform distributions.

\smallskip
\noindent\textbf{Computing environment.}
All experiments were conducted on a dedicated GPU server equipped with four NVIDIA A100 GPUs (80GB HBM2e memory per device), dual AMD EPYC 7763 CPUs, and 1 TB of system RAM. This configuration enabled efficient large-scale natural language processing and statistical modeling. The software environment was based on Python 3.10, with pandas and numpy supporting data preprocessing and aggregation, spaCy and textstat enabling linguistic feature extraction, and scikit-learn providing implementations for TF–IDF vectorization, non-negative matrix factorization, and similarity measures. Statistical modeling, including interrupted time-series regressions, was implemented with statsmodels, while visualizations were produced using matplotlib and seaborn with customized formatting for heatmaps, ridge plots, and temporal trend figures.

\subsection{(RQ1) Submission Volumes and Growth Rates}
\label{sec:RQ1}

\begin{figure}[!t]
  \centering
  % Subfigure 1
  \subfigure[Monthly counts with 3-month moving average\label{fig:rq1_monthly}]{
    \includegraphics[width=0.95\linewidth]{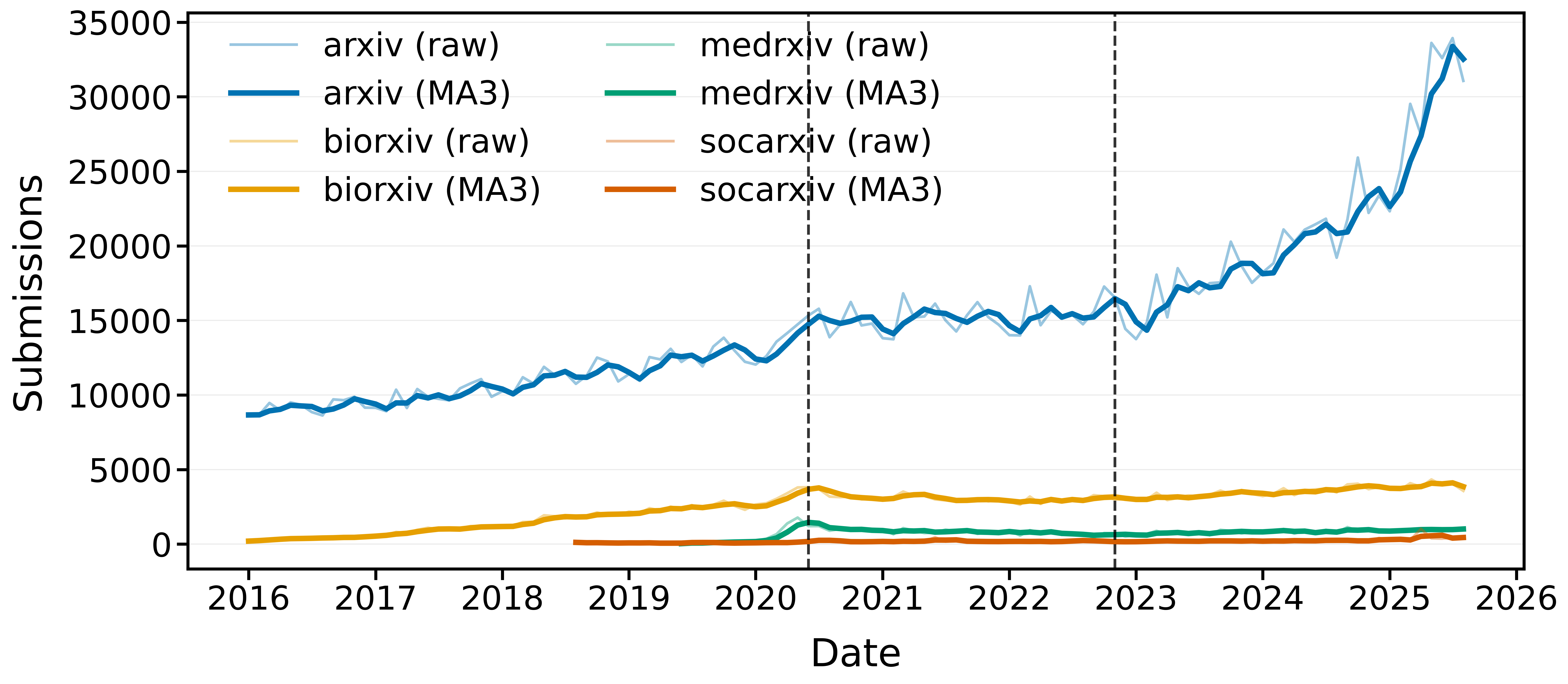}
  }

  % Subfigure 2
  \subfigure[Cumulative submission totals\label{fig:rq1_cumulative}]{
    \includegraphics[width=0.95\linewidth]{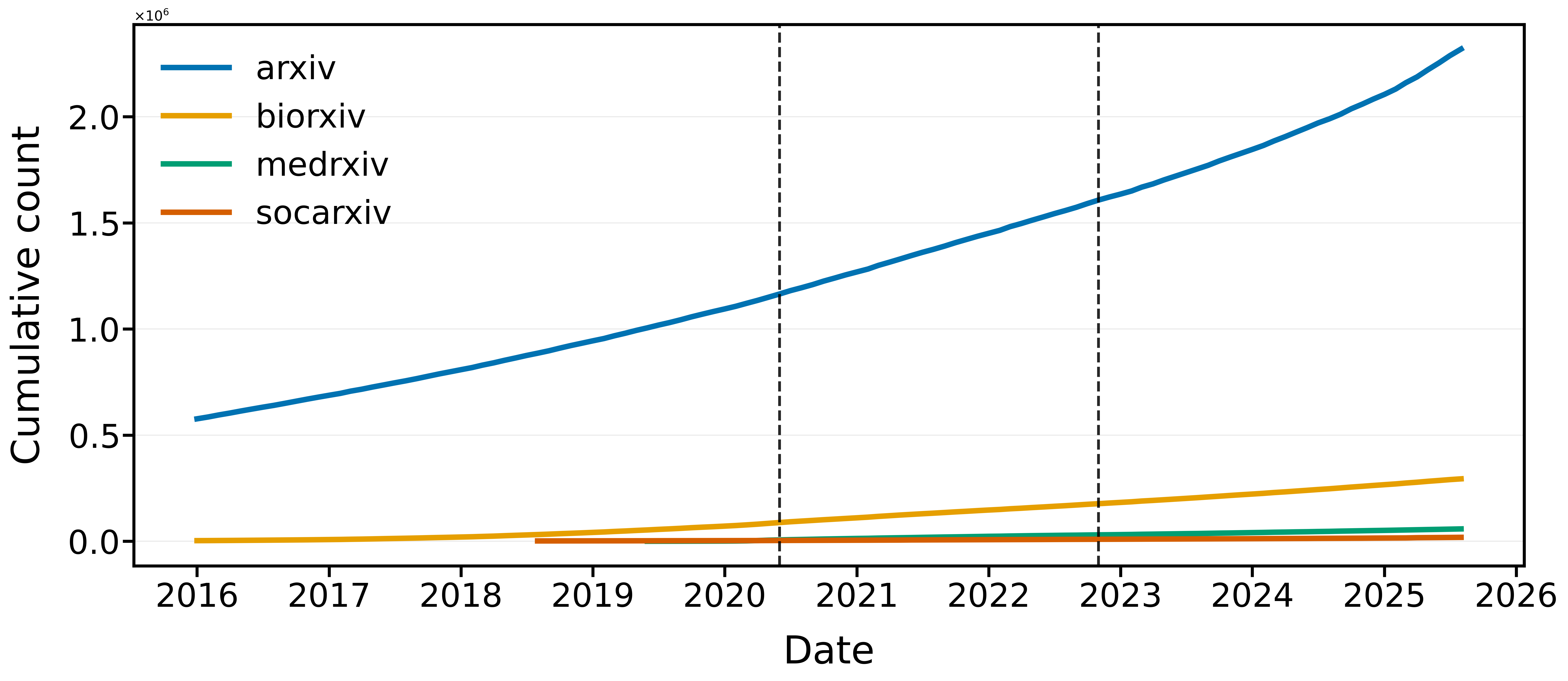}
  }

  % Subfigure 3
  \subfigure[Year-over-Year growth rates\label{fig:rq1_yoy}]{
    \includegraphics[width=0.95\linewidth]{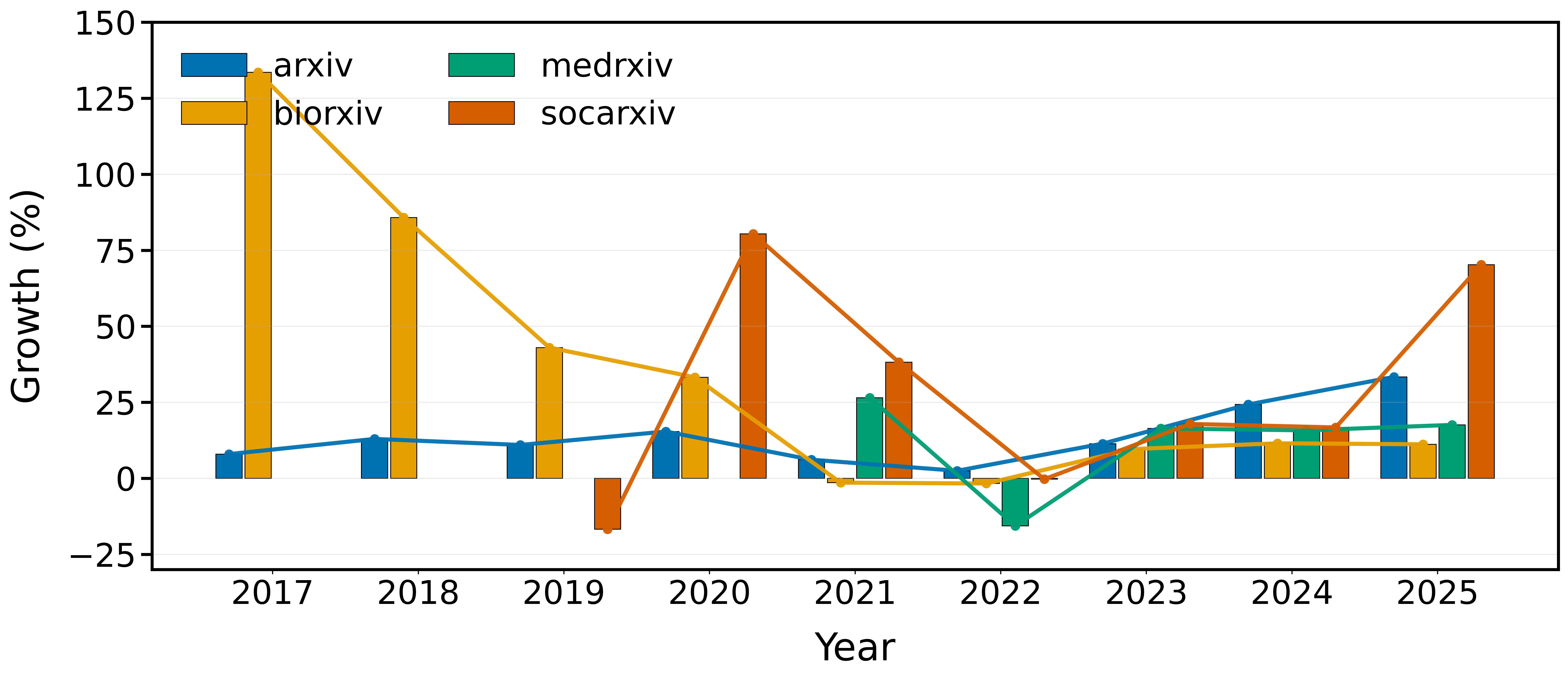}
  }

  \caption{Submission across repositories.  
  The plots summarize monthly submission counts with a 3-month moving average, cumulative totals, and year-over-year growth rates. Vertical markers denote the GPT-3 release (June 2020) and the ChatGPT release (November 2022).} %这三个图可以加上黑色frame 然后图里字都放大些 包括后面的几个数据图也是放大字体 
  \label{fig:rq1_overview}
\end{figure}

\smallskip
\noindent\textbf{a. Monthly counts with MA(3).}
\textcolor{teal}{Fig.\ref{fig:rq1_monthly}} plots monthly submission counts for arXiv, bioRxiv, medRxiv, and SocArXiv from 2016 to 2025, overlaid with a three-month moving average to attenuate short-run volatility. Vertical markers at June 2020 and November 2022 indicate the GPT-3 API release and the public launch of ChatGPT, respectively. Across repositories, the smoothed trajectories reveal steady pre-2020 growth, a pronounced pandemic-era bulge that is most visible in medRxiv, and a post-2022 re-acceleration that is clearest in arXiv and bioRxiv. The arXiv series shows a long upward trend with seasonal undulations; following the ChatGPT marker, the MA(3) line exhibits a sustained increase in slope rather than an abrupt step change. bioRxiv displays a similar but smaller pattern: growth stabilizes after the early pandemic spike and begins to climb more quickly after late 2022. medRxiv records a sharp transient surge during 2020 consistent with COVID-19 activity, returns toward its pre-surge baseline, and shows modest but persistent gains following November 2022 once short-term outliers are smoothed by MA(3). SocArXiv, with lower absolute volumes and greater month-to-month noise, nonetheless trends upward over the study window with a gradual post-2022 rise.

% These visual patterns are consistent with the interrupted time-series estimates in Table \ref{tab:rq1-its}, which suggest limited evidence for a positive level jump at the ChatGPT marker but indicate statistically significant increases in post-marker slopes for arXiv and bioRxiv, and smaller positive slope changes for medRxiv and SocArXiv. Taken together, the MA(3) curves and the regression results provide convergent evidence that the introduction of widely accessible LLMs coincides with an acceleration in submission dynamics, after accounting for seasonality and the pandemic perturbation.

\begin{table}[!t]
  \centering
  \caption{Interrupted time series for monthly submissions. 
  Coefficients are reported with HAC-robust standard errors.}
  \label{tab:rq1-its}
  \resizebox{\linewidth}{!}{
  \begin{tabular}{l|rrrr}
    \toprule
    Repository & Level shift (GPT-3) & Slope $\Delta$ (GPT-3) 
    & Level shift (ChatGPT) & Slope $\Delta$ (ChatGPT) \\
    \midrule
    \midrule
    arXiv    & \inc{3,879.23} \textcolor{gray}{\scriptsize (1,268.07)} 
             & \dec{53.17} \textcolor{gray}{\scriptsize (19.55)}  
             & \dec{42,045.06} \textcolor{gray}{\scriptsize (6,592.83)} 
             & \inc{477.38} \textcolor{gray}{\scriptsize (70.32)} \\
    bioRxiv  & \inc{2,870.01} \textcolor{gray}{\scriptsize (432.44)}   
             & \dec{56.79} \textcolor{gray}{\scriptsize (6.73)}   
             & \dec{2,348.77} \textcolor{gray}{\scriptsize (399.08)}   
             & \inc{29.60} \textcolor{gray}{\scriptsize (5.61)}   \\
    medRxiv  & \inc{1,215.02} \textcolor{gray}{\scriptsize (336.74)}   
             & \dec{121.57} \textcolor{gray}{\scriptsize (35.13)} 
             & \dec{714.57} \textcolor{gray}{\scriptsize (300.84)}     
             & \inc{18.04} \textcolor{gray}{\scriptsize (7.87)}   \\
    SocArXiv & \dec{415.91} \textcolor{gray}{\scriptsize (235.03)}     
             & \inc{22.80} \textcolor{gray}{\scriptsize (10.78)}   
             & \dec{234.19} \textcolor{gray}{\scriptsize (239.99)}     
             & \inc{2.43} \textcolor{gray}{\scriptsize (4.59)}    \\
    \bottomrule
  \end{tabular}}
\end{table}

\smallskip
\noindent\textbf{b. Cumulative submissions.}  
\textcolor{teal}{Fig.\ref{fig:rq1_cumulative}} traces the cumulative growth of preprint submissions across the four repositories. It reveals a broadly monotonic upward pattern, yet the pace of accumulation differs markedly across platforms. \textit{arXiv} dominates in scale, surpassing two million submissions by 2025, and displays a steady expansion without evident stagnation. \textit{bioRxiv} shows a strong acceleration phase between 2018 and 2021, followed by a more gradual slope thereafter, suggesting partial saturation of its growth trajectory. \textit{medRxiv}, beginning later in 2019, exhibits the steepest early incline, reflecting rapid adoption during the COVID-19 period before settling into a more stable accumulation pattern. \textit{SocArXiv}, though smaller in scale, demonstrates consistent linear growth without the inflection points observed in the biomedical repositories.

\smallskip
\noindent\textbf{c. Year-over-Year growth.}  
\textcolor{teal}{Fig.\ref{fig:rq1_yoy}} presents the year-over-year (YoY) growth rates of preprint submissions, thereby normalizing for repository scale and highlighting acceleration or deceleration dynamics. The results reveal sharp contrasts across platforms. \textit{medRxiv} records the most dramatic spike in 2020, consistent with the surge of pandemic-related manuscripts, before reverting to a more moderate trajectory. By contrast, \textit{bioRxiv} displays sustained but gradually declining growth rates after peaking in the late 2010s, indicative of a maturing platform. \textit{arXiv} shows relative stability, with moderate fluctuations but without extreme volatility, suggesting a more entrenched user base. Finally, \textit{SocArXiv} exhibits irregular yet consistently positive growth, reflective of its smaller scale and emergent disciplinary community.  

\smallskip
\noindent\textbf{d. ITS regression model.} 
\textcolor{teal}{Table~\ref{tab:rq1-its}} reports the interrupted time-series regression estimates for monthly submissions, with coefficients interpreted as immediate level shifts and slope changes following the releases of GPT-3 and ChatGPT. The results indicate that \textit{arXiv} experienced a pronounced positive level shift after GPT-3 (+3,879 submissions per month) but a subsequent decline in growth rate (–53 per month). In contrast, the release of ChatGPT corresponded to a large negative level effect (–42,045) followed by a strong positive slope (+477), suggesting a short-term contraction but accelerated recovery thereafter. \textit{bioRxiv} exhibits a similar though more moderate pattern, with GPT-3 linked to a positive level change (+2,870) but declining slope (–56), and ChatGPT associated with a negative level effect (–2,348) but positive slope (+29.6). For \textit{medRxiv}, GPT-3 corresponds to a small positive level and sharp negative slope (–121), while ChatGPT effects are comparatively modest (–715; +18). Finally, \textit{SocArXiv} displays attenuated responses, with weakly negative level changes and near-zero slope adjustments, consistent with its smaller and more heterogeneous scale.

These parameter estimates align with the descriptive evidence presented in \textcolor{teal}{Fig.\ref{fig:rq1_overview}}. The monthly submission trajectories (\textcolor{teal}{Fig.\ref{fig:rq1_monthly}}) visually capture the abrupt shifts and subsequent slope changes that the ITS model quantifies, particularly the 2022–2023 fluctuations in arXiv. The cumulative totals (\textcolor{teal}{Fig.\ref{fig:rq1_cumulative}}) reveal the longer-term implications of these slope adjustments, with arXiv and bioRxiv exhibiting accelerated accumulation after ChatGPT, whereas medRxiv and SocArXiv remain relatively flat. Year-over-year growth rates (\textcolor{teal}{Fig.\ref{fig:rq1_yoy}}) further corroborate the slope estimates, showing widespread deceleration following GPT-3 and renewed acceleration after ChatGPT, most strongly for arXiv and bioRxiv.

\begin{insightbox}
\textbf{Finding 1:} Submission dynamics exhibit a pattern of short-term disruption followed by medium-term acceleration. ITS estimates show significant post-ChatGPT slope increases in all repositories, most notably in \textit{arXiv} (\(\textbf{+477}\), \(p<0.01\)) and \textit{bioRxiv} (\(+29.6\), \(p<0.05\)).
\end{insightbox}

\subsection{(RQ2) Collaboration Patterns and Author Productivity}
\label{sec:RQ2}

\begin{figure}[!t]
  \centering
  \includegraphics[width=\linewidth]{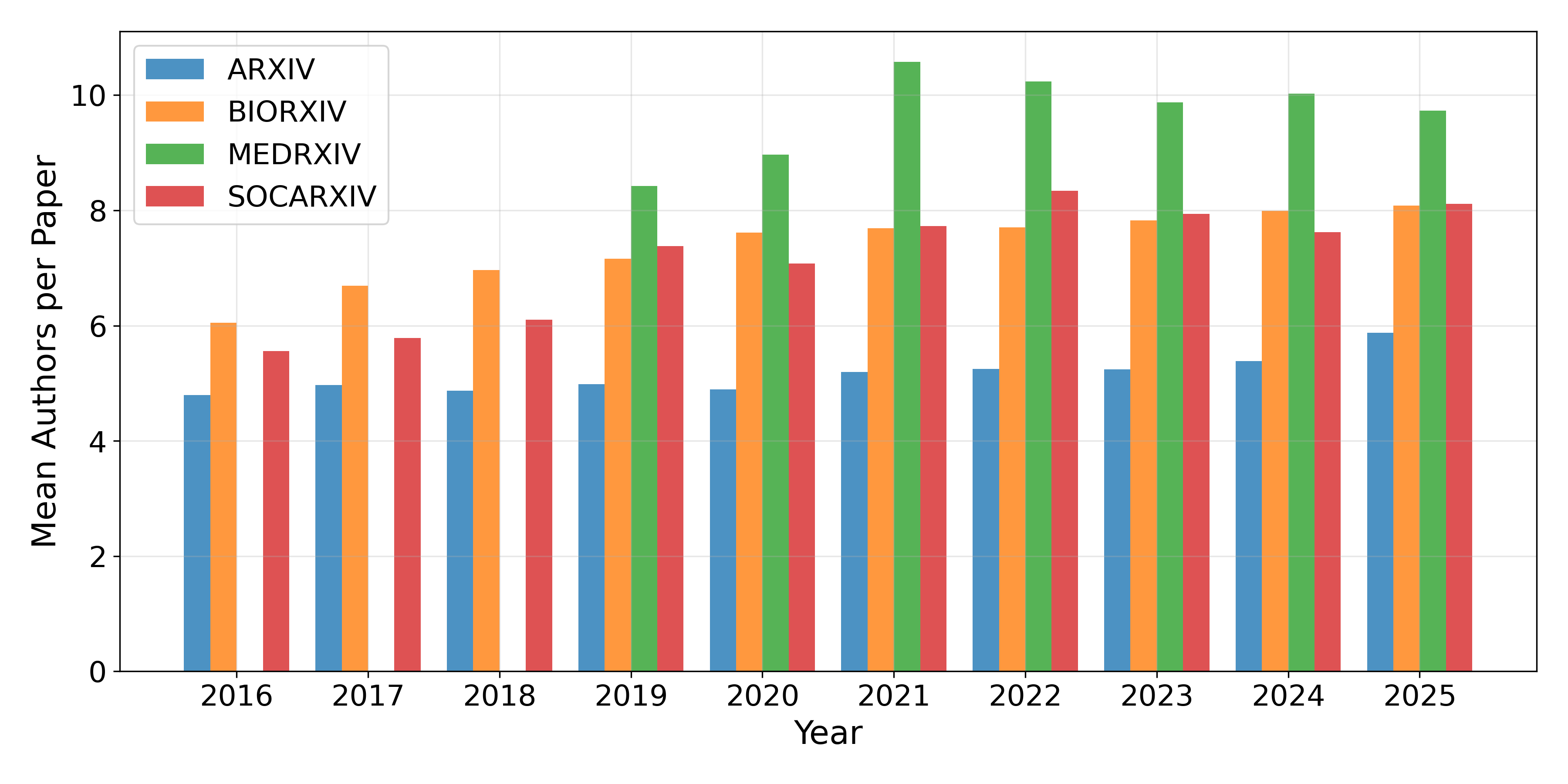}
  \caption{Mean number of authors per paper.}
  \label{fig:rq2_mean_authors}
\end{figure}

\smallskip
\noindent\textbf{a. Mean authors.}  
\textcolor{teal}{Fig.\ref{fig:rq2_mean_authors}} depicts the mean number of authors per paper across four repositories from 2016 to 2025. The overall trend reveals a steady increase in collaboration intensity, with notable differences across domains. In \textit{arXiv}, growth is relatively modest, rising from approximately 4.8 authors per paper in 2016 to just under 6.0 by 2025, suggesting that collaboration in the physical sciences and mathematics has expanded but at a measured pace. In contrast, \textit{bioRxiv} exhibits a more substantial upward trajectory, from around 6.0 to over 8.0 authors on average, reflecting the increasingly team-based nature of life science research. The steepest rise occurs in \textit{medRxiv}, which begins with 8.4 authors per paper in 2019 and surpasses 10 by 2021, stabilizing at this elevated level in subsequent years. This indicates that clinical and biomedical research relies heavily on large-scale, multi-institutional collaborations, particularly during and after the COVID-19 pandemic. \textit{SocArXiv} shows an intermediate pattern, increasing from about 5.5 authors to above 8.0, suggesting gradual but clear movement toward larger collaborative groups in the social sciences.

\begin{figure}[!t]
  \centering
  \includegraphics[width=\linewidth]{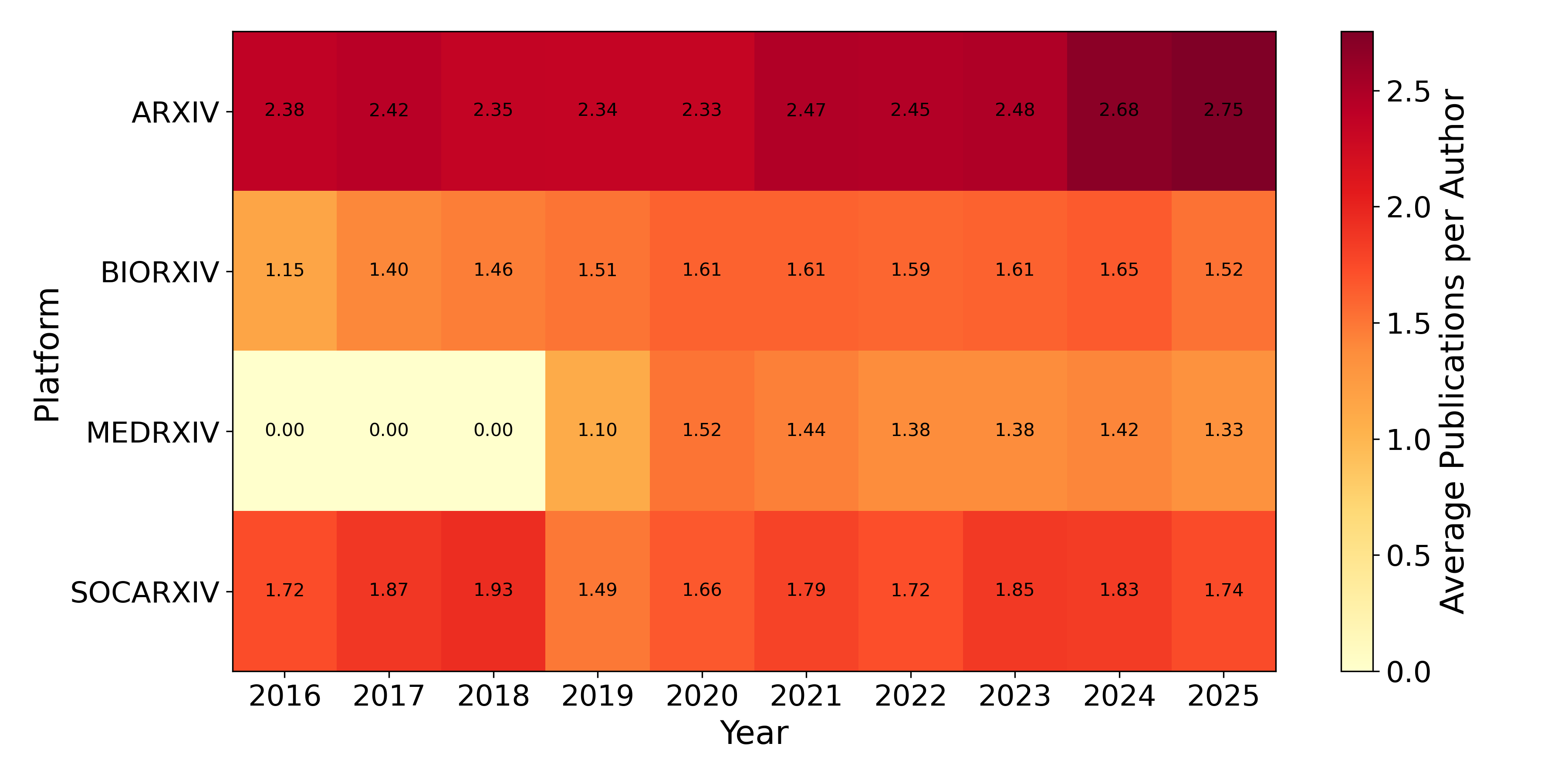}
  \caption{Annual publications per author.}
  \label{fig:rq2_step2}
\end{figure}

\smallskip
\noindent\textbf{b. Author productivity.}
To complement the collaboration analysis, we next examine author-level productivity, defined as the average number of publications per author per year. \textcolor{teal}{Fig.\ref{fig:rq2_step2}} visualizes these values as a heatmap across repositories from 2016 to 2025. The results indicate distinct productivity regimes: authors on \textit{arXiv} consistently publish more frequently, averaging between 2.3 and 2.7 papers annually, with a gradual upward drift in the post-ChatGPT period. By contrast, \textit{bioRxiv} and \textit{SocArXiv} show moderate productivity levels, typically in the range of 1.4–1.9 papers per year, with modest gains over time but without a clear structural break. \textit{medRxiv}, which entered the ecosystem only in 2019, initially experienced a sharp productivity surge associated with the COVID-19 pandemic but stabilized thereafter at roughly 1.3–1.5 papers per year.

\begin{figure}[!t]
  \centering
  \subfigure[Distribution of revision intervals]{%
    \includegraphics[width=\linewidth]{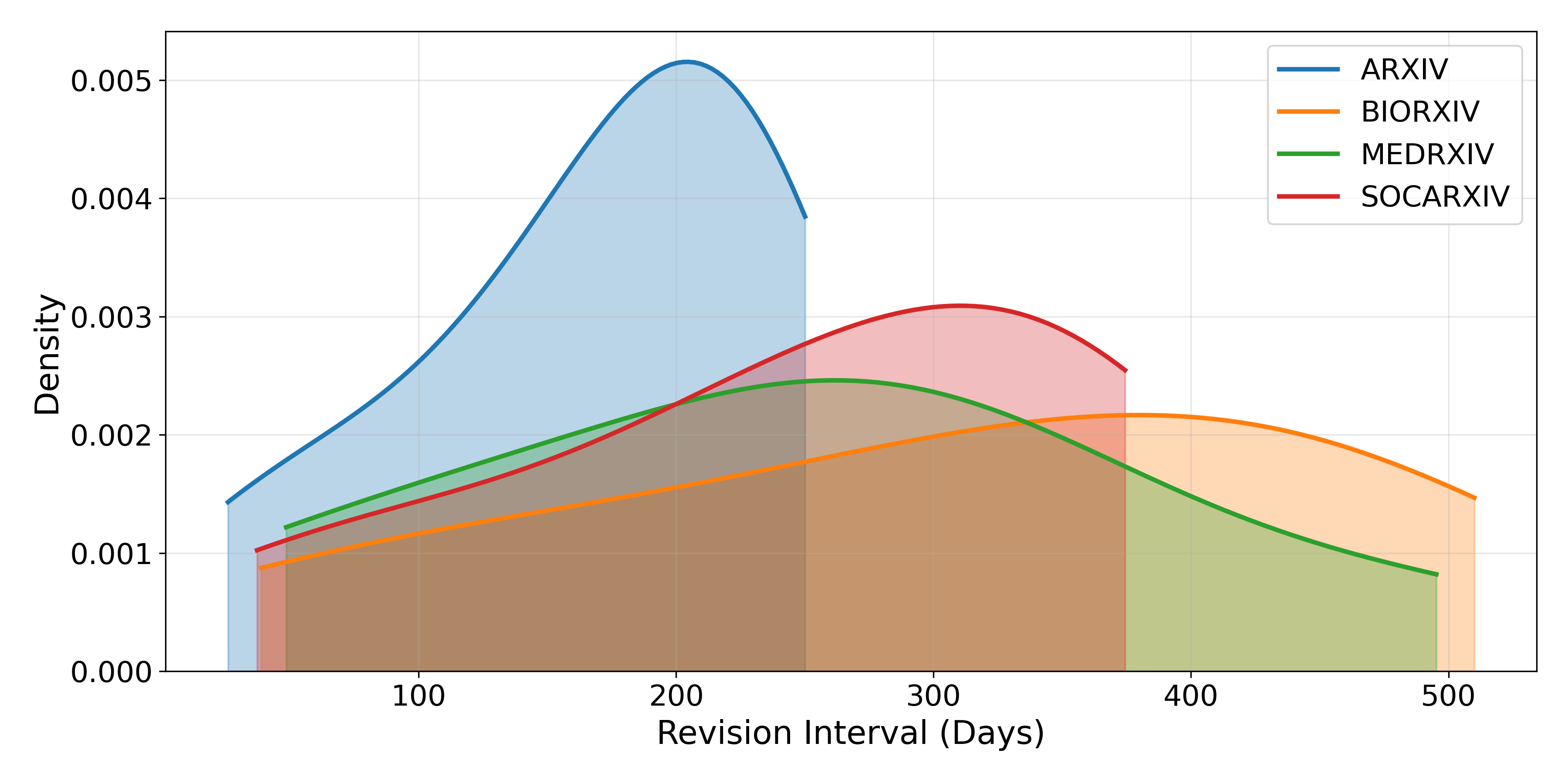}%
    \label{fig:rq2_step3_density}}
  \vskip\baselineskip
  \subfigure[Average revision interval with LLM milestones]{%
    \includegraphics[width=\linewidth]{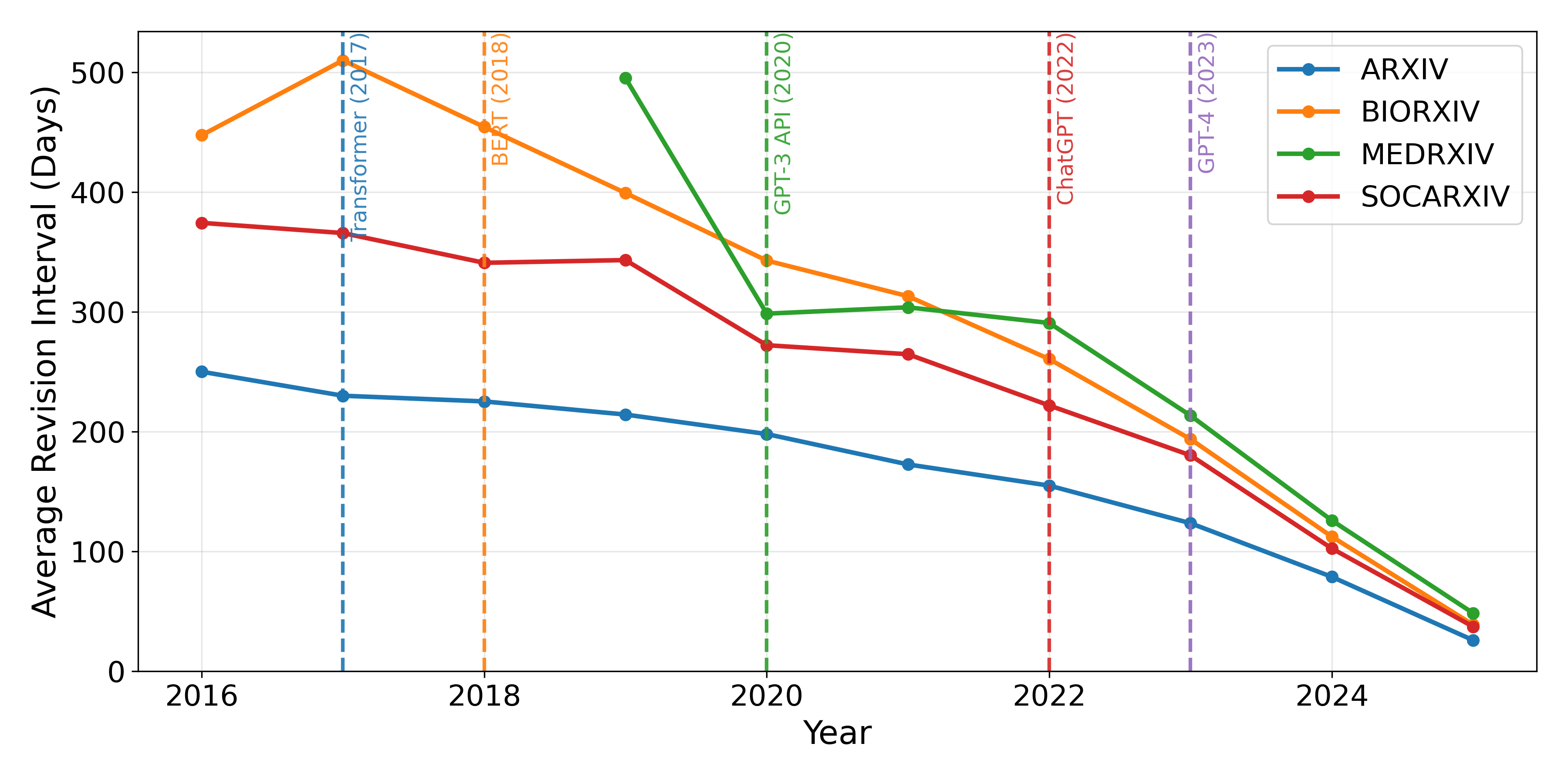}%
    \label{fig:rq2_step3_line}}
  \caption{Revision tempo across repositories. (a) Density estimates show the distribution of revision intervals. (b) Yearly averages reveal declining revision lags, with annotated vertical markers indicating major LLM milestones (Transformer, BERT, GPT-3 API, ChatGPT, GPT-4).}
  \label{fig:rq2_step3}
\end{figure}

\smallskip
\noindent\textbf{c. Revision tempo.} 
\textcolor{teal}{Fig.\ref{fig:rq2_step3_density}} presents complementary views of revision tempo. Panel (a) displays kernel density estimates of revision intervals, highlighting systematic disciplinary differences. \textit{arXiv} is characterized by a sharp mode around 200 days, reflecting the rapid turnover typical of physics and computer science. In contrast, \textit{bioRxiv} and \textit{SocArXiv} exhibit broader and heavier-tailed distributions, suggesting slower and more heterogeneous revision cycles. \textit{medRxiv} occupies an intermediate position, with evidence of pandemic-era acceleration.

Panel (b) extends this analysis by plotting annual averages of revision intervals from 2016 to 2025, annotated with vertical markers for major LLM milestones. A pronounced downward trend is evident across all repositories, with particularly steep declines after the release of GPT-3 and ChatGPT. This convergence suggests that faster iteration cycles are not merely discipline-specific but may also reflect systemic pressures introduced by the widespread availability of generative AI tools. Nonetheless, cross-repository differences persist, with \textit{arXiv} consistently operating on the shortest cycles and the social sciences retaining comparatively longer lags.

\begin{insightbox}
\textbf{Finding 2:} Mean team sizes expanded across all repositories (e.g., from 4.8→6.0 on \textit{arXiv}, 6.0→8.0 on \textit{bioRxiv}), while individual productivity remained heterogeneous ($\approx$2.5 vs.\ 1.5 papers per author per year). Revision intervals shortened by roughly \textbf{30\%} post-2020, indicating that LLM-enabled workflows accelerated dissemination consistently across domains.
\end{insightbox}

\begin{figure}[!h]
  \centering
  \subfigure[Flesch–Kincaid Grade Level (FKGL)]{%
    \includegraphics[width=\linewidth]{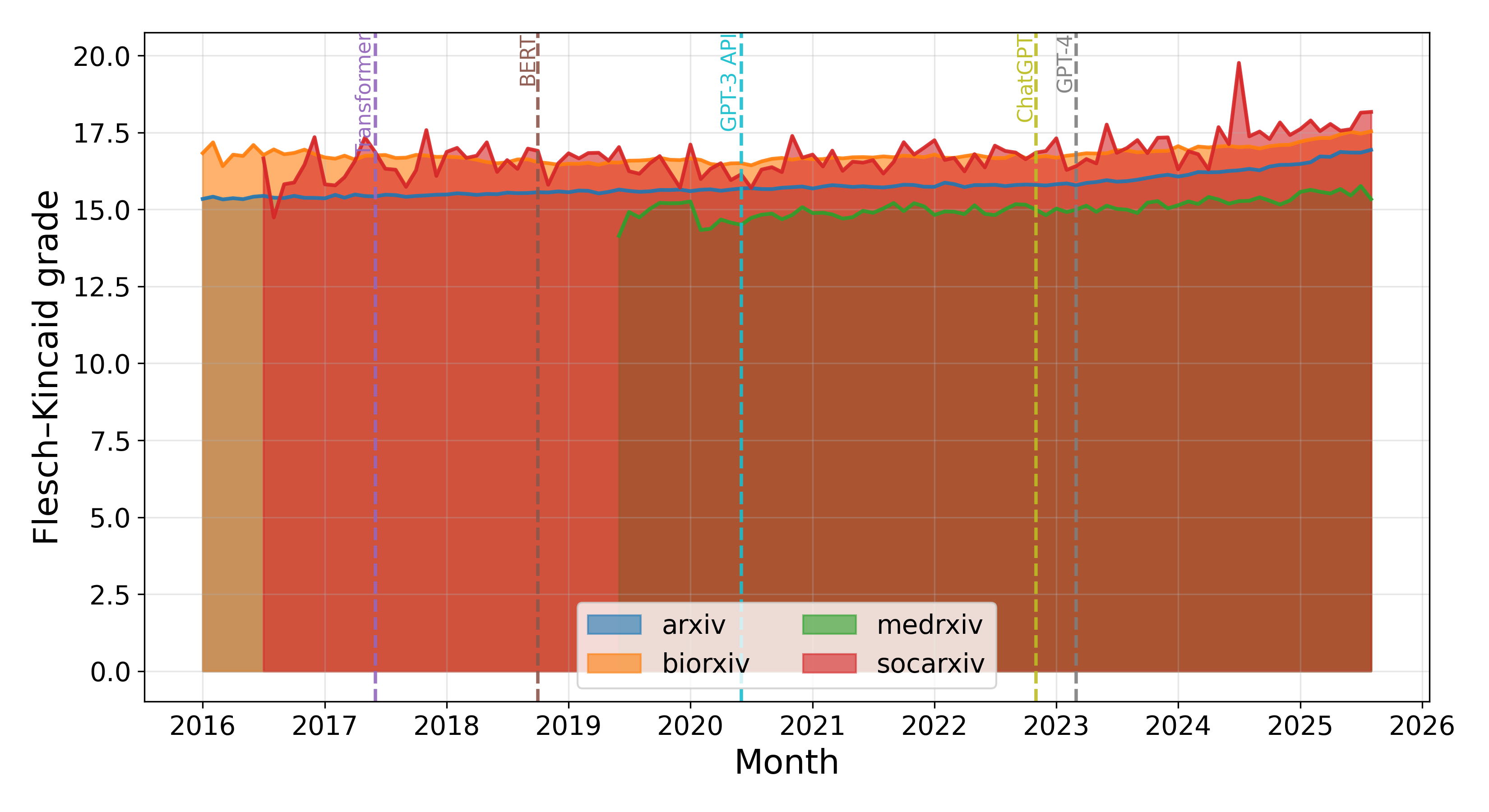}%
    \label{fig:fkgl_4sites}}
  \vskip\baselineskip
  \subfigure[Flesch Reading Ease (FRE)]{%
    \includegraphics[width=\linewidth]{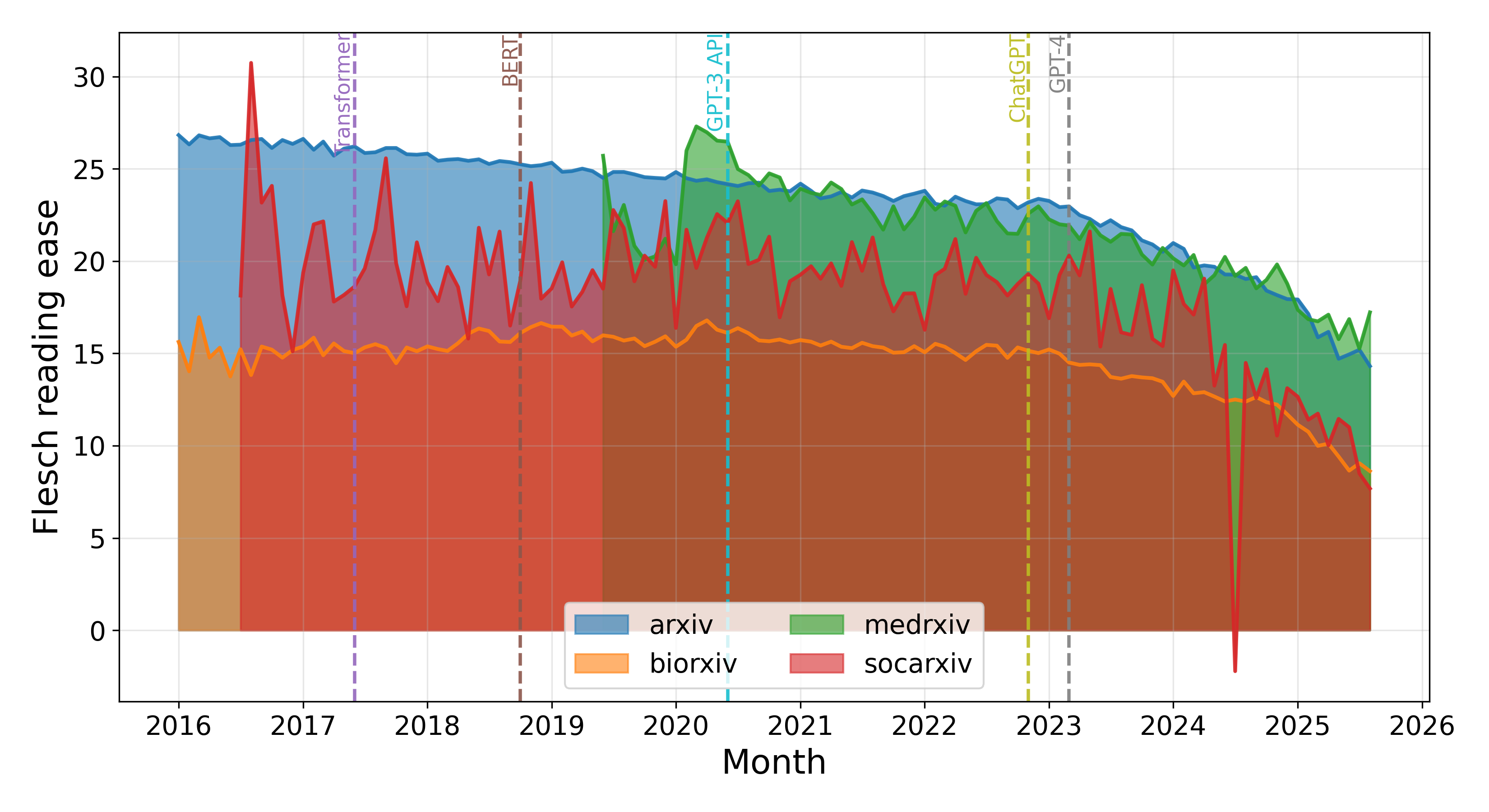}%
    \label{fig:fre_4sites}}
  \caption{Readability analysis across repositories.}
  \label{fig:rq3_readability}
\end{figure}

\subsection{(RQ3) Readability and Linguistic Complexity of Abstracts}
\label{sec:RQ3}

\smallskip
\noindent\textbf{a. Readability analysis.}  
\textcolor{teal}{Fig.\ref{fig:fkgl_4sites}} reports the Flesch–Kincaid Grade Level (FKGL), while \textcolor{teal}{Fig.\ref{fig:fre_4sites}} presents the Flesch Reading Ease (FRE). Both indicators reveal a gradual shift in readability across repositories. On arXiv and bioRxiv, FKGL values remain consistently high (between 15–18), suggesting that abstracts typically require graduate-level proficiency to comprehend. However, after late 2022, modest upward drifts are observed, particularly in SocArXiv, where FKGL values show spikes exceeding 18, consistent with a denser and more formal style of writing. In contrast, medRxiv maintains comparatively lower grade levels, reflecting domain-specific conventions that favor accessibility in medical communication.

The FRE scores provide a complementary perspective. Across all four platforms, values are clustered in the 10–30 range, which corresponds to texts that are challenging to read for a general audience. A noticeable downward trend is evident from 2022 onward, with scores declining by 2–3 points on average, particularly in arXiv and medRxiv. This indicates that preprints have become progressively harder to read, likely due to increasing lexical density and syntactic elaboration.

\begin{figure*}[!t]
  \centering
  \subfigure[Average sentence length]{%
    \includegraphics[width=0.24\textwidth]{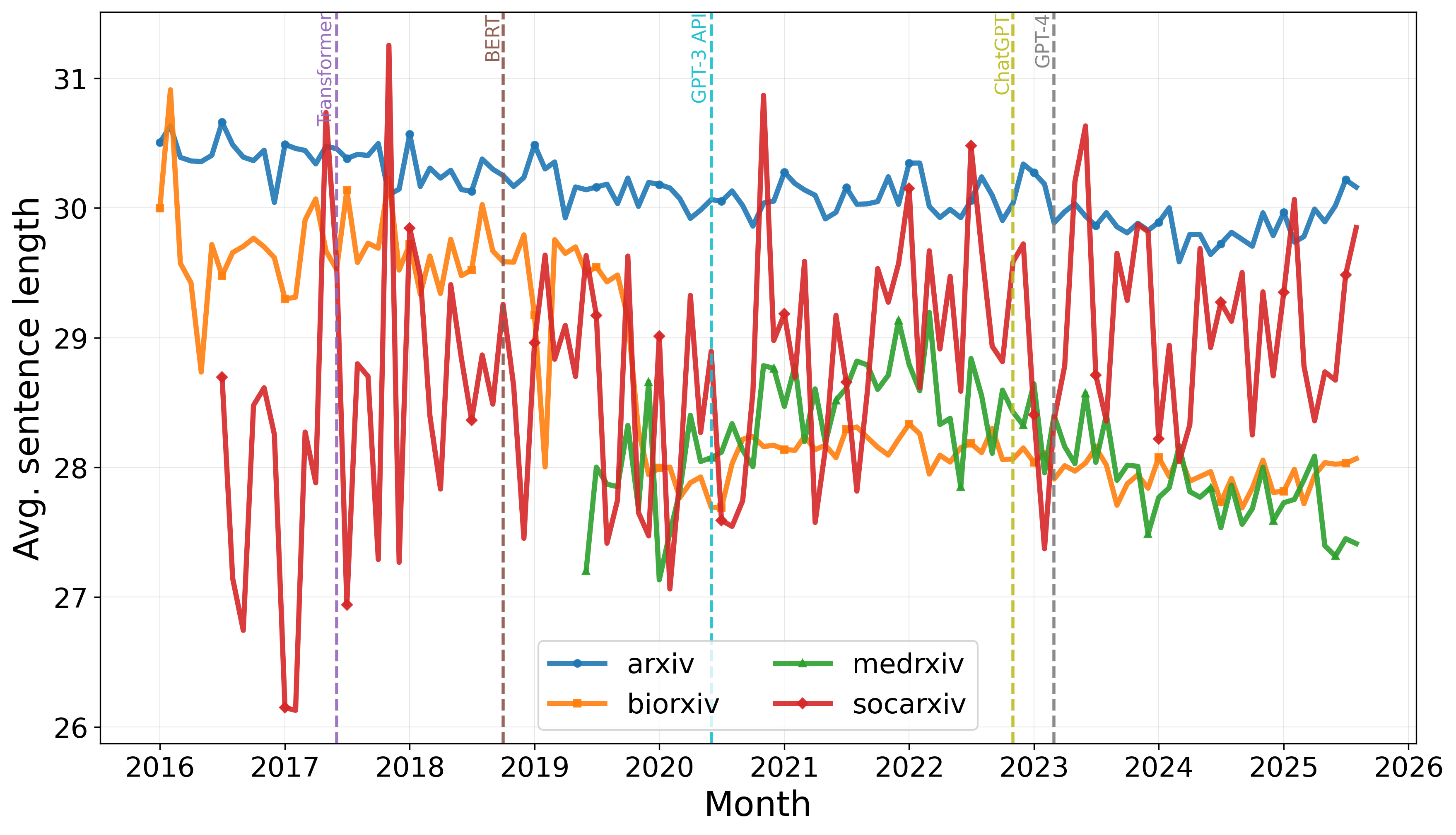}%
    \label{fig:sentlen}}
  \hfill
  \subfigure[Difficult\textendash word ratio]{%
    \includegraphics[width=0.24\textwidth]{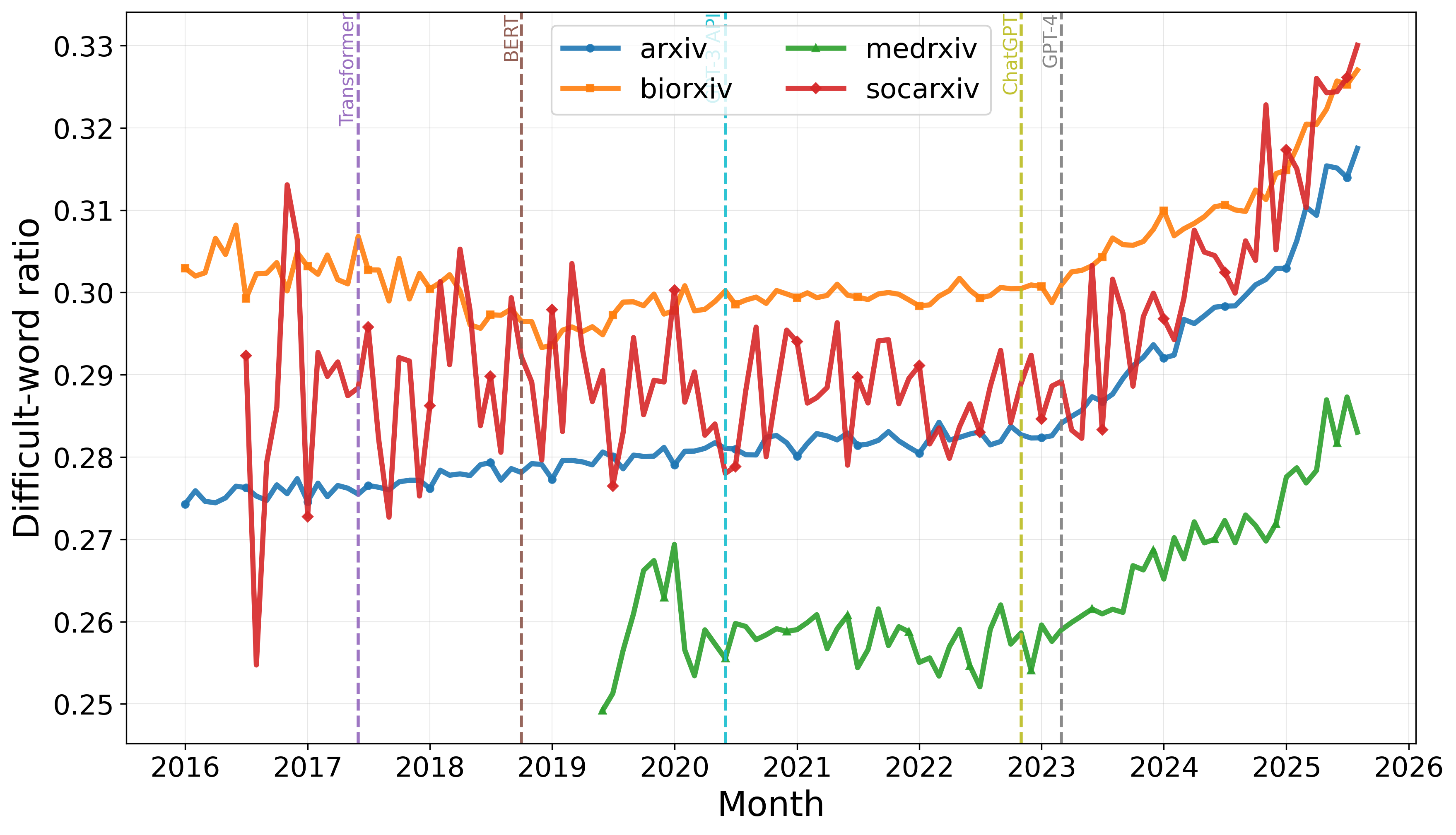}%
    \label{fig:difficult}}
  \hfill
  \subfigure[Lexical density]{%
    \includegraphics[width=0.24\textwidth]{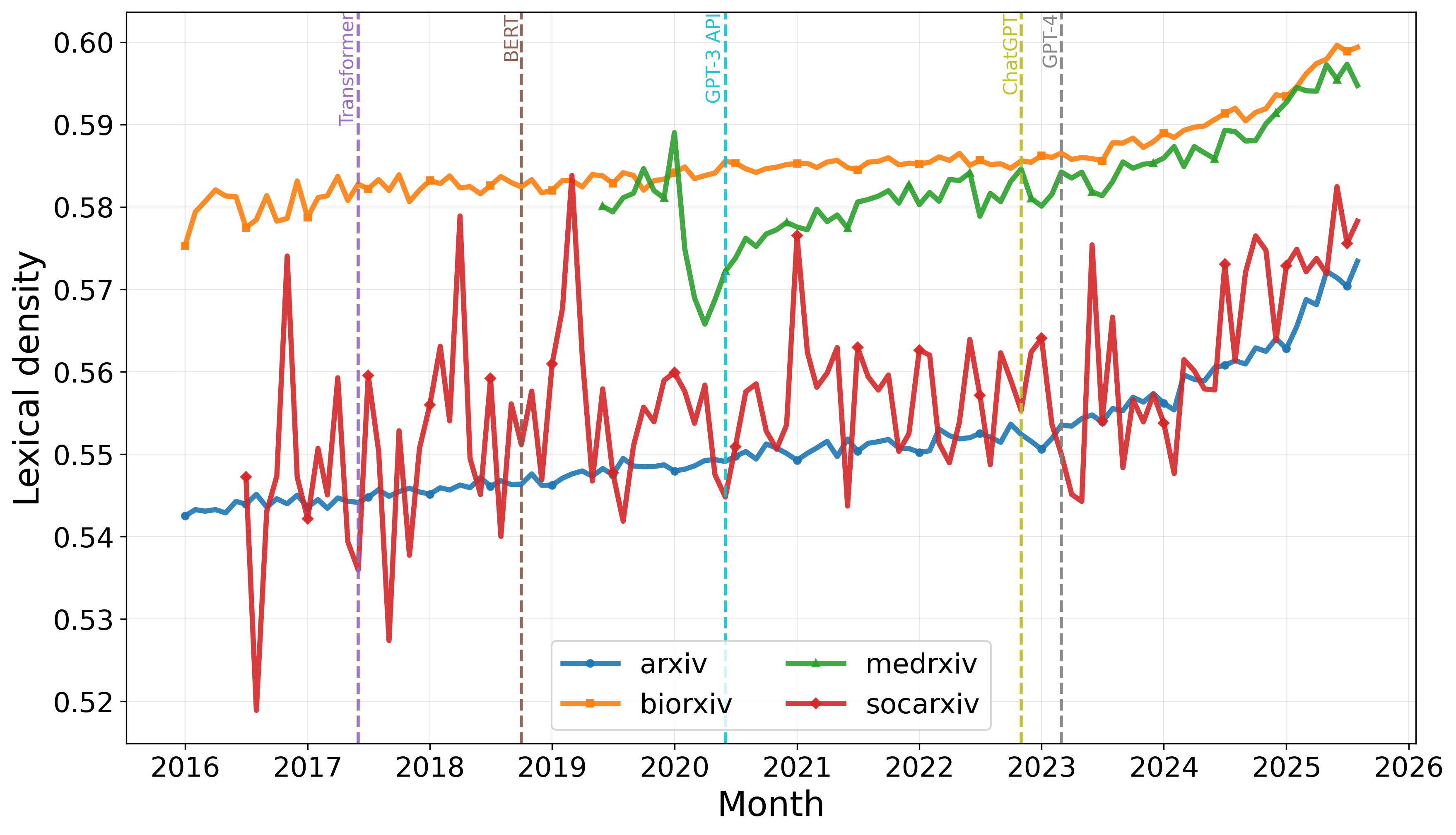}%
    \label{fig:lexical}}
  \hfill
  \subfigure[Subordination ratio]{%
    \includegraphics[width=0.24\textwidth]{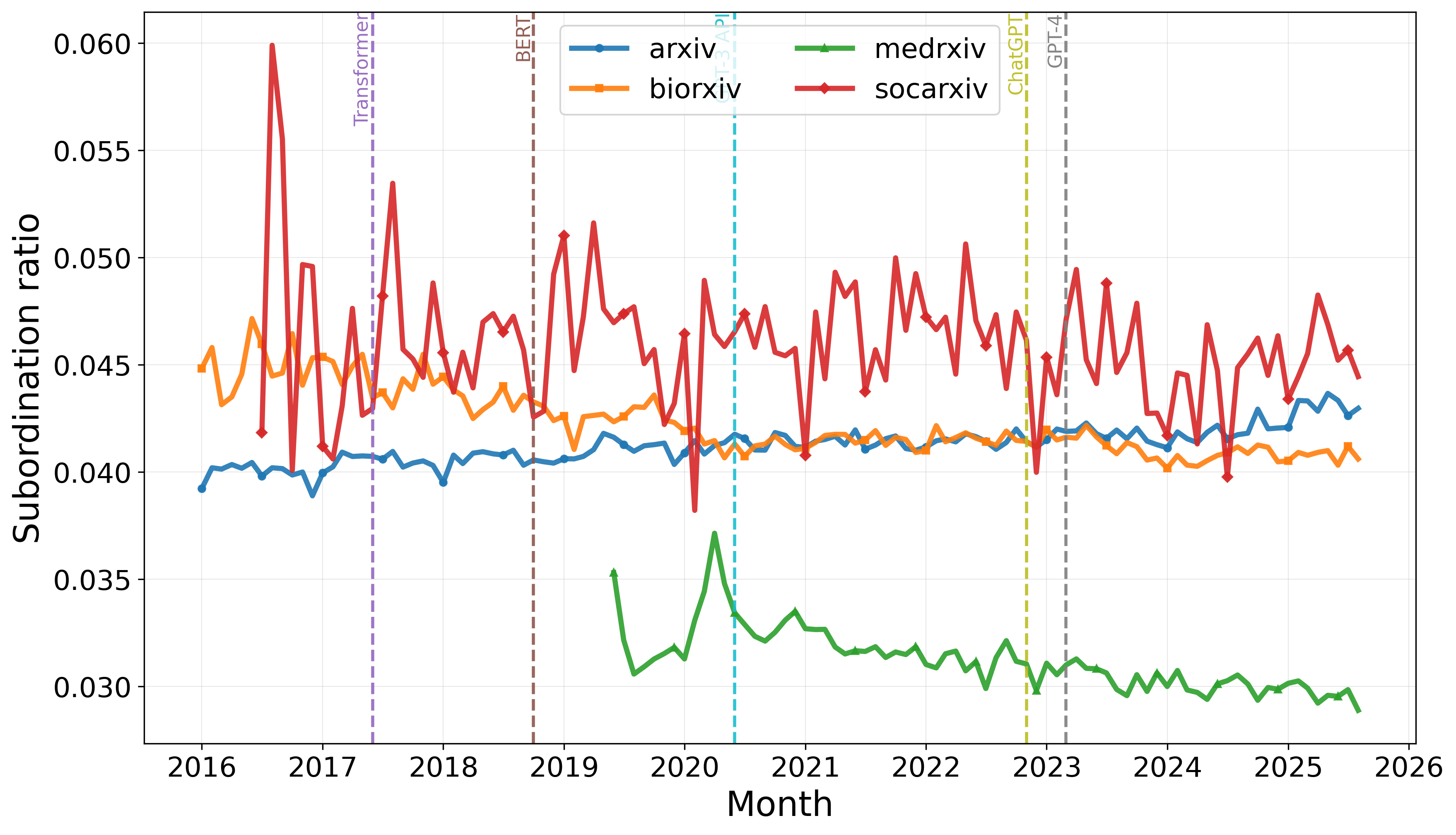}%
    \label{fig:subordination}}
  \caption{Linguistic complexity indicators across repositories (2016–2025): (a) average sentence length, (b) difficult\textendash word ratio, (c) lexical density, and (d) subordination ratio. Vertical dashed lines mark major LLM milestones.}
  \label{fig:rq3_complexity}
\end{figure*}

\begin{figure*}[!t]
  \centering
  \subfigure[arXiv]{%
    \includegraphics[width=0.24\textwidth]{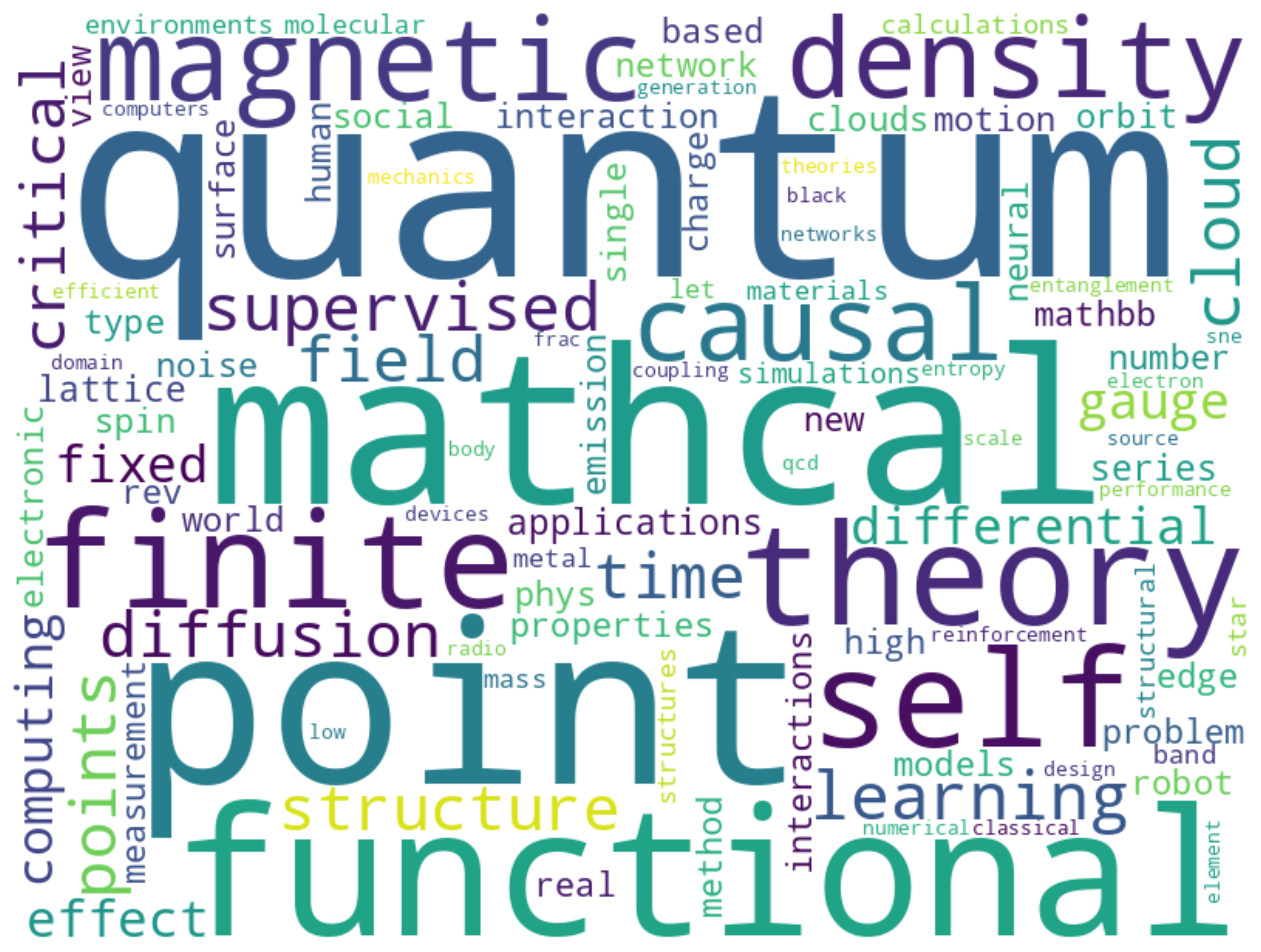}%
    \label{fig:arxiv_wc}}
  \hfill
  \subfigure[bioRxiv]{%
    \includegraphics[width=0.24\textwidth]{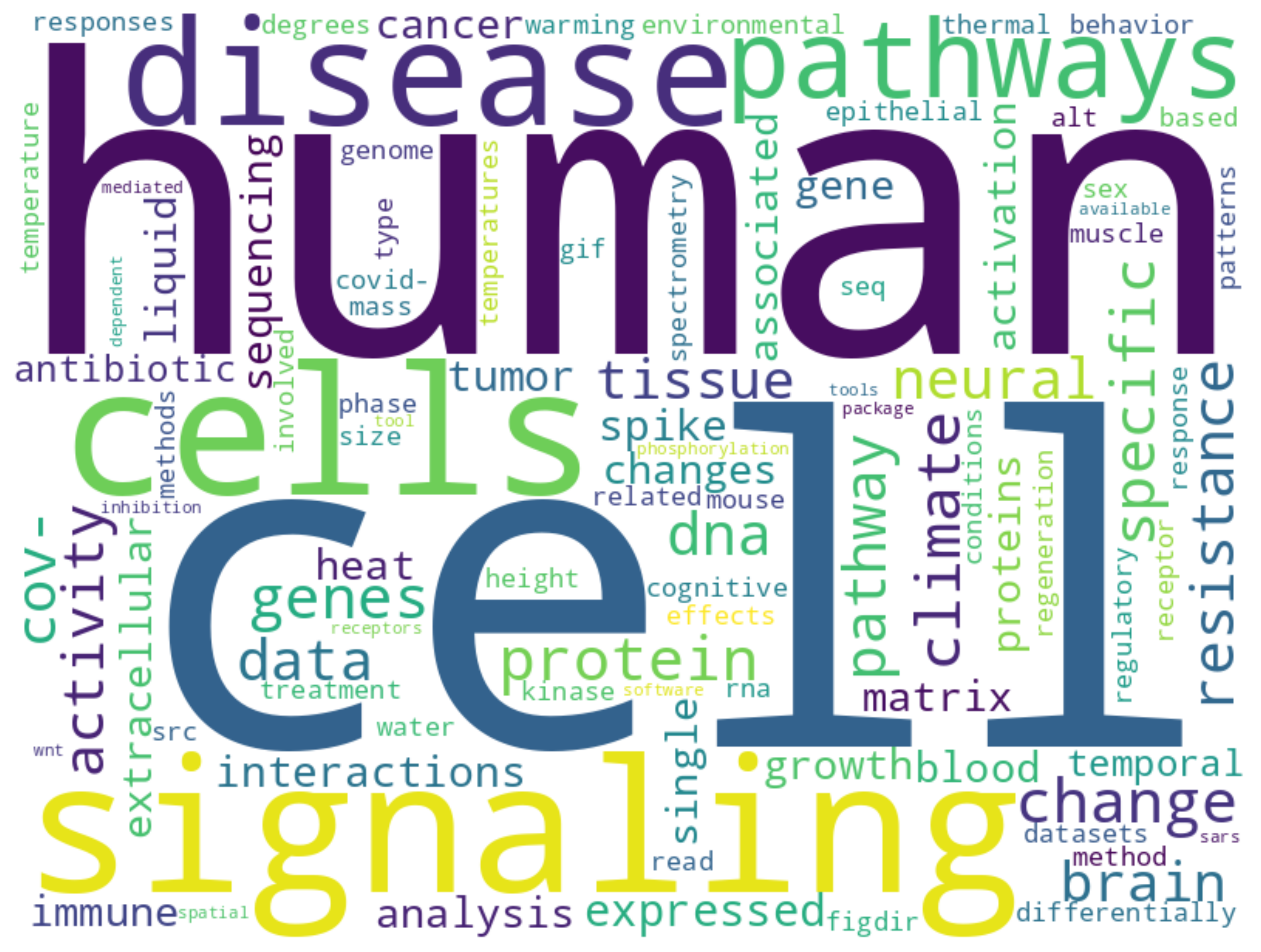}%
    \label{fig:biorxiv_wc}}
  \hfill
  \subfigure[medRxiv]{%
    \includegraphics[width=0.24\textwidth]{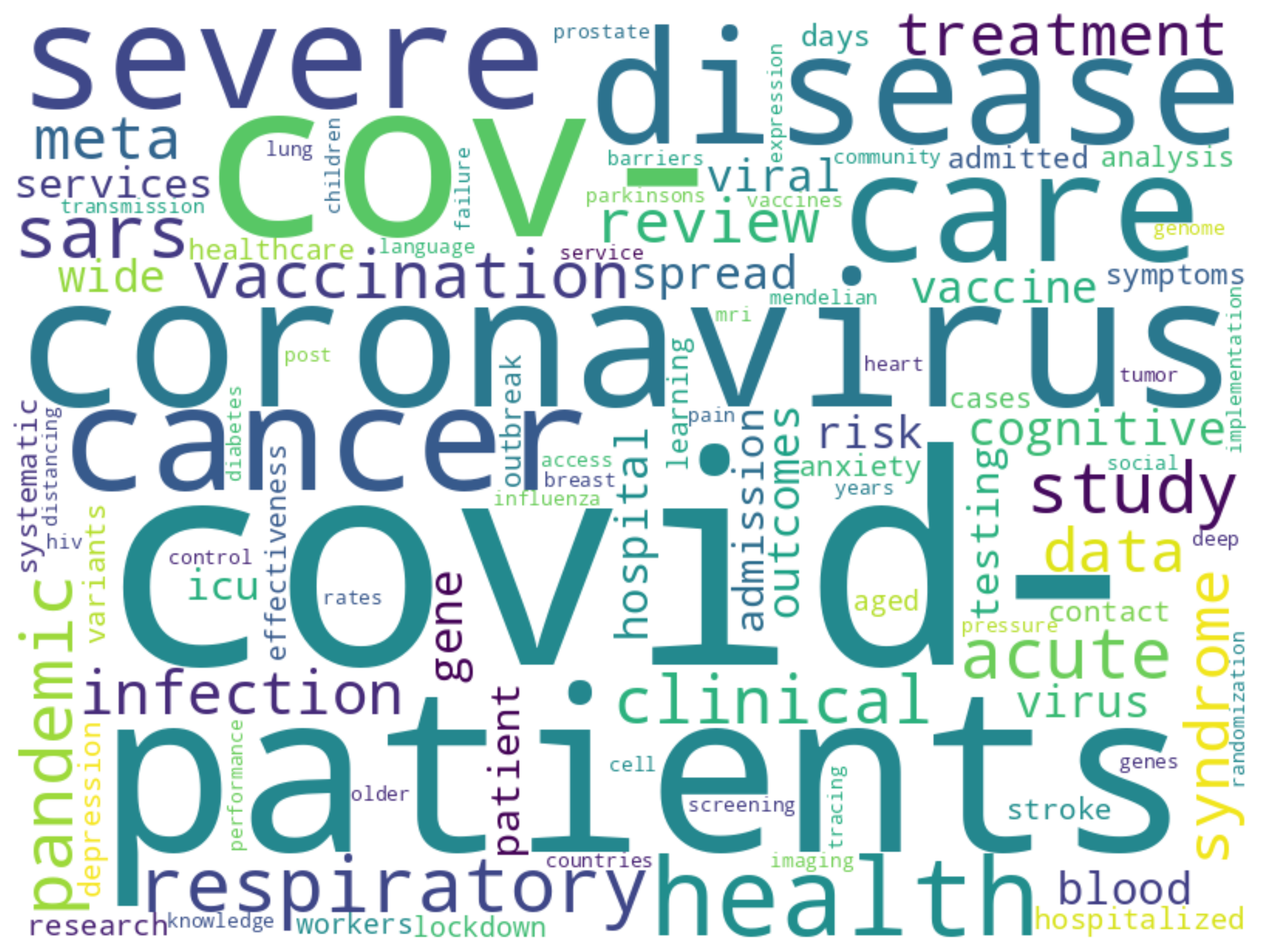}%
    \label{fig:medrxiv_wc}}
  \hfill
  \subfigure[SocArXiv]{%
    \includegraphics[width=0.24\textwidth]{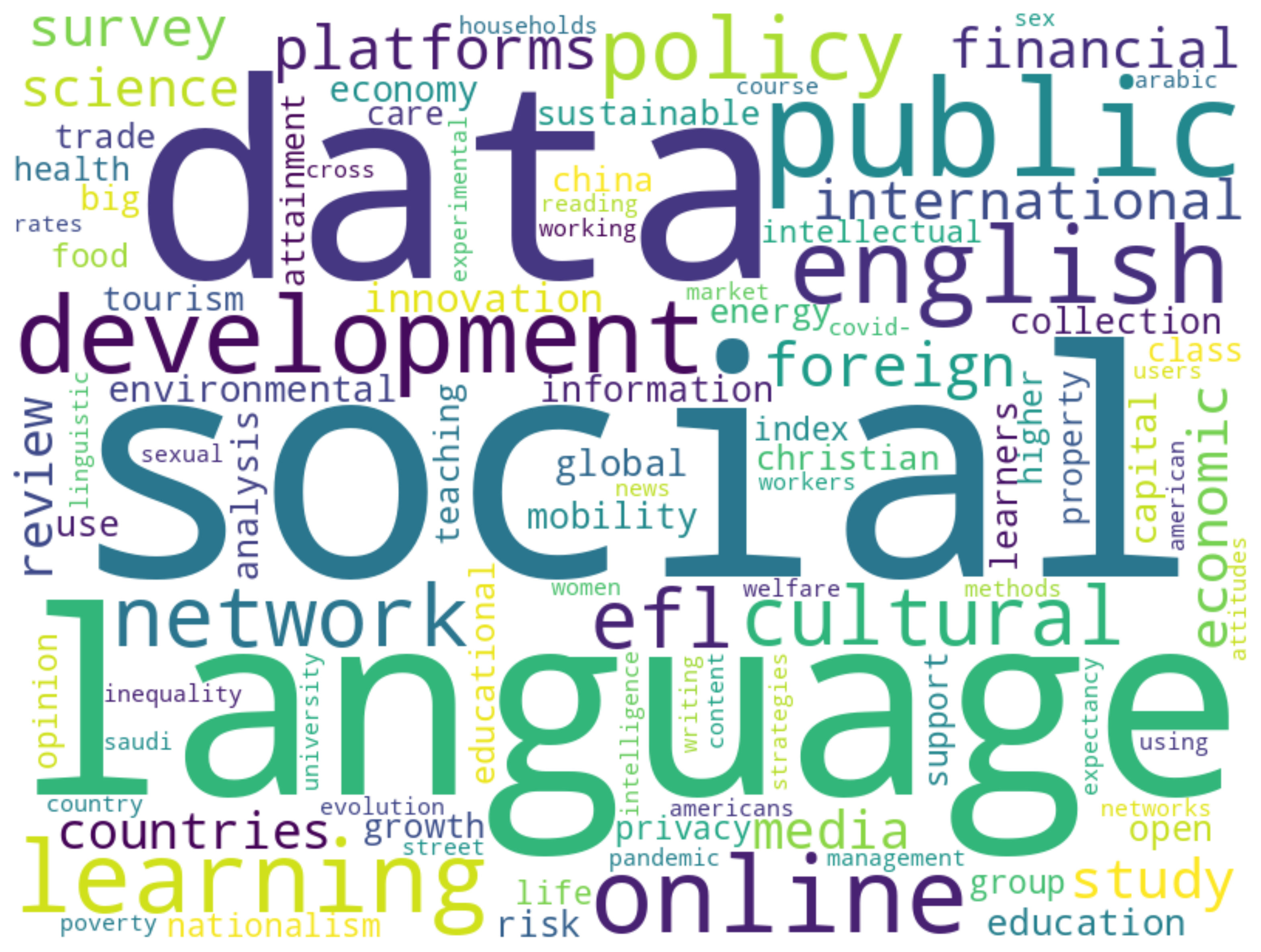}%
    \label{fig:socarxiv_wc}}
  \caption{Representative AI-related topic extracted from NMF topic modeling across repositories. Word clouds highlight domain-specific thematic orientations, with arXiv dominated by computational and mathematical terms, bioRxiv by genomic and cellular processes, medRxiv by clinical and pandemic-related themes, and SocArXiv by social and policy-oriented discourse.}
  \label{fig:rq5_wordclouds}
\end{figure*}
\smallskip
\noindent\textbf{b. Linguistic complexity.} The four linguistic complexity indicators jointly highlight both stability and subtle shifts in textual practices across repositories. Average sentence length (\textcolor{teal}{Fig.\ref{fig:sentlen}}) remains largely stable throughout 2016--2025, with \textit{arXiv} consistently producing longer sentences of approximately 30 words, while \textit{bioRxiv}, \textit{medRxiv}, and \textit{SocArXiv} show slightly shorter structures. This persistence indicates that syntactic elaboration has not undergone fundamental transformations, even after the release of ChatGPT. In contrast, the difficult-word ratio (\textcolor{teal}{Fig.\ref{fig:difficult}}) reveals modest upward trends, most evident in \textit{arXiv} and \textit{SocArXiv}, with \textit{bioRxiv} maintaining the highest overall levels. 

Lexical density (\textcolor{teal}{Fig.\ref{fig:lexical}}) and subordination ratio (\textcolor{teal}{Fig.\ref{fig:subordination}}) provide further insight into informational richness and syntactic depth. Lexical density converges across repositories, with values stabilizing in the 0.55--0.60 range, indicating consistently high proportions of content-bearing words. The small upward drift after 2022 reflects incremental increases in informational load rather than abrupt stylistic changes. Subordination ratios, by contrast, highlight persistent cross-platform differences: \textit{arXiv} and \textit{bioRxiv} remain stable at around 0.04, \textit{medRxiv} tends lower near 0.03, while \textit{SocArXiv} exhibits more volatility with episodic spikes.

\begin{figure*}[!t]
  \centering
  \subfigure[arXiv subject distributions]{%
    \includegraphics[width=0.24\textwidth]{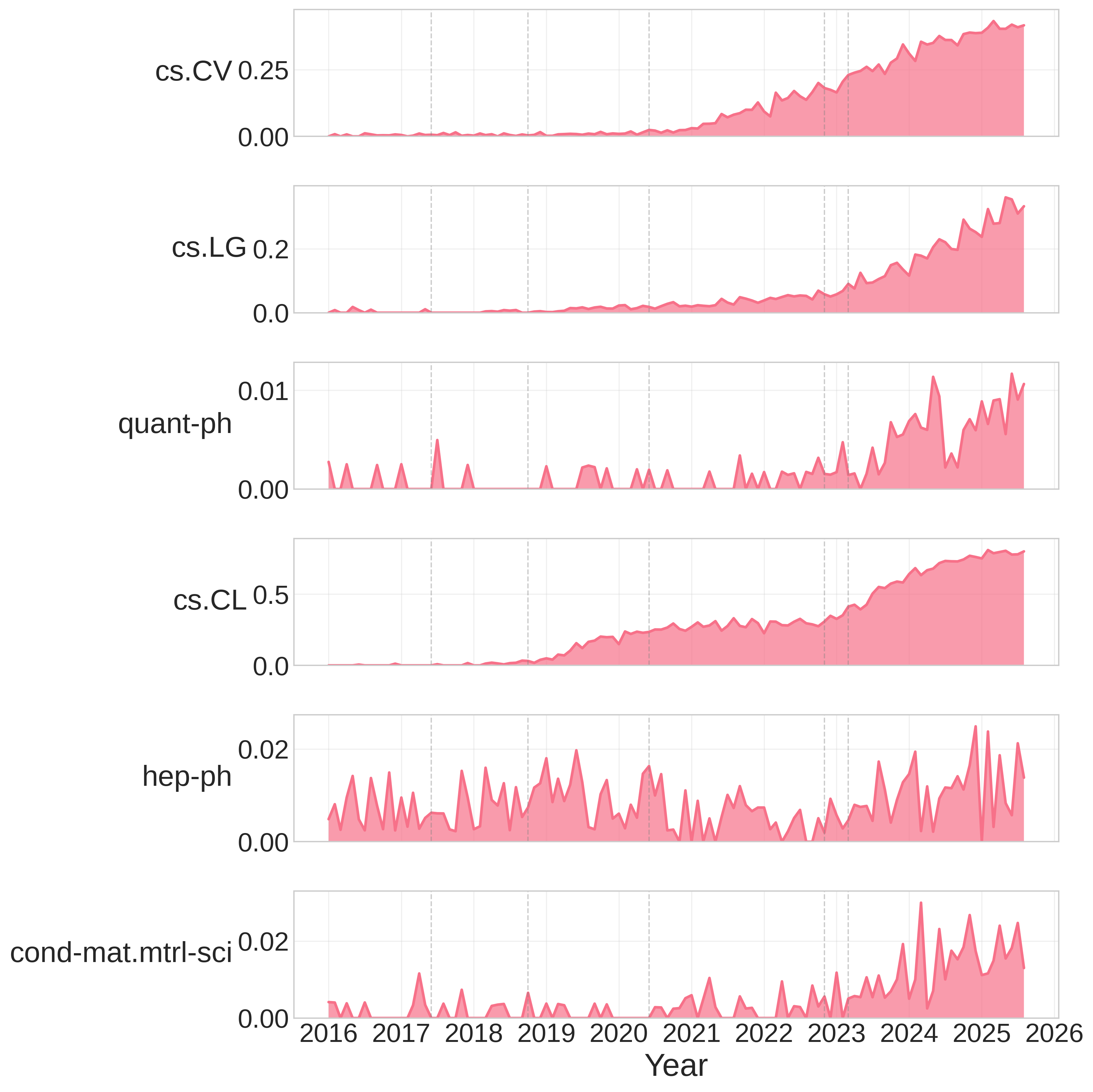}%
    \label{fig:arxiv_ridge}}
  \hfill
  \subfigure[bioRxiv subject distributions]{%
    \includegraphics[width=0.24\textwidth]{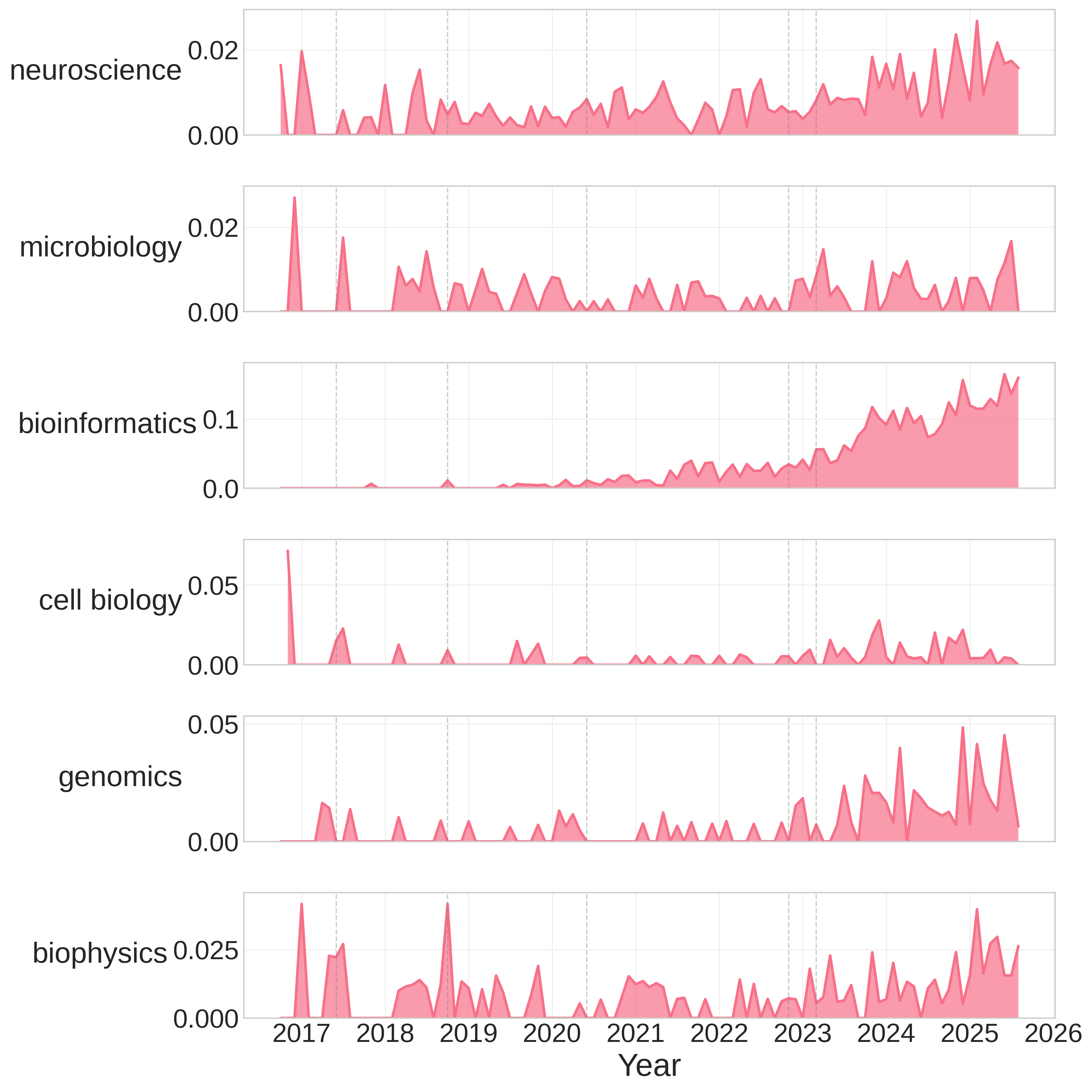}%
    \label{fig:biorxiv_ridge}}
  \hfill
  \subfigure[medRxiv subject distributions]{%
    \includegraphics[width=0.24\textwidth]{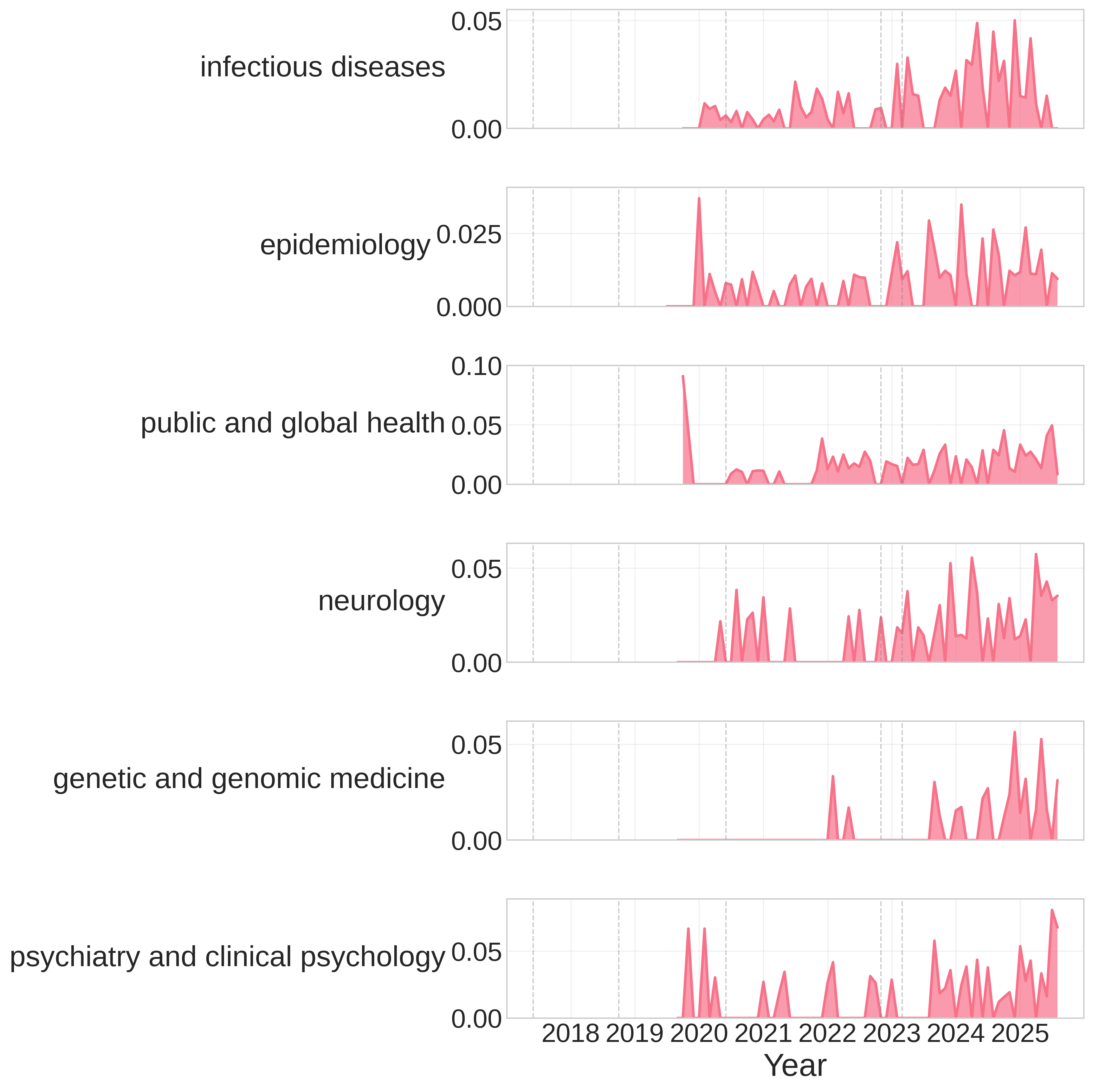}%
    \label{fig:medrxiv_ridge}}
  \hfill
  \subfigure[SocArXiv subject distributions]{%
    \includegraphics[width=0.24\textwidth]{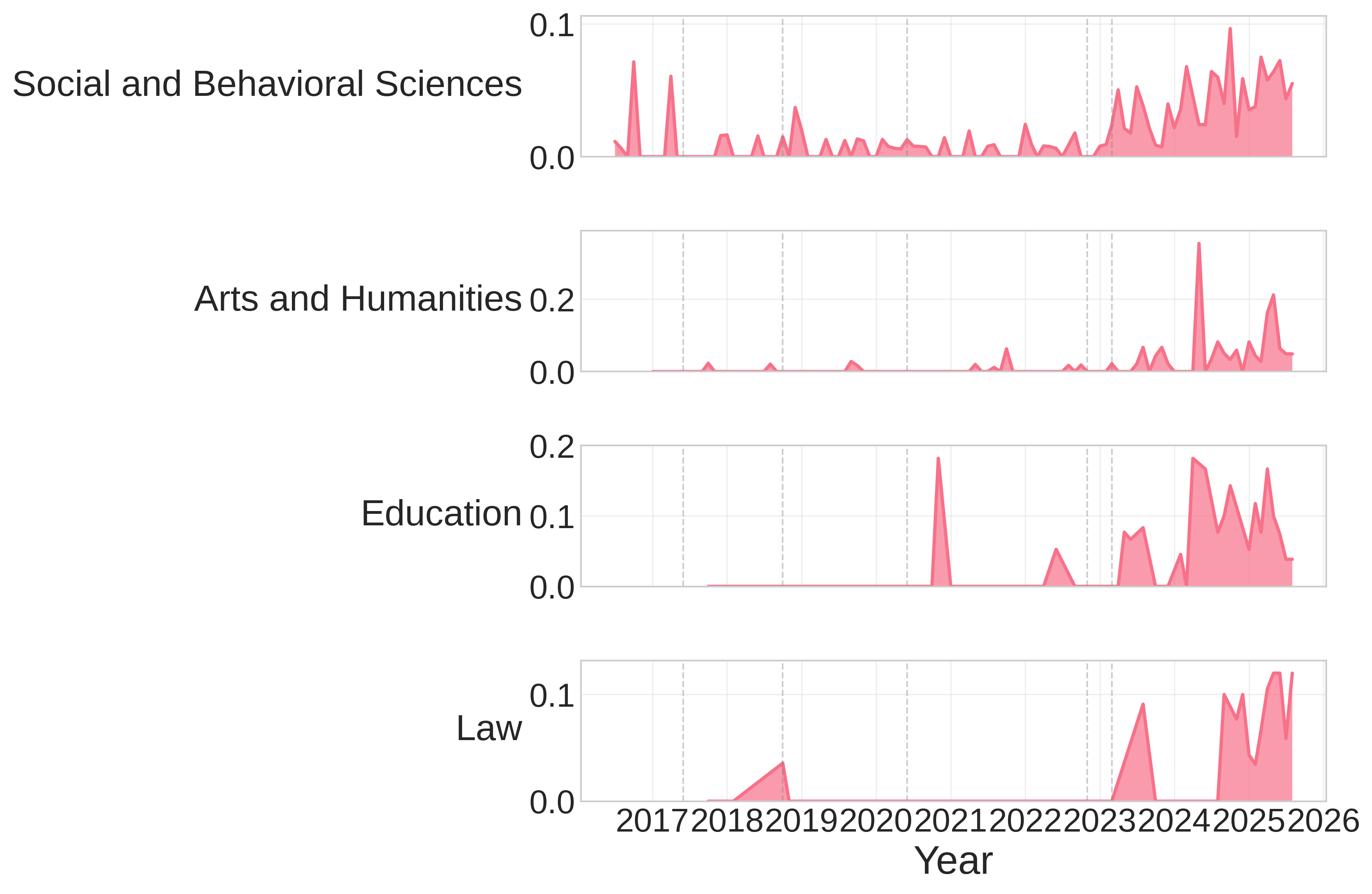}%
    \label{fig:socarxiv_ridge}}
  \caption{Disciplinary distributions of AI-related document shares across repositories (2016–2025). Each panel visualizes the temporal dynamics of AI adoption within subject-specific categories, highlighting both early-adopting and lagging fields.}
  \label{fig:rq7_ridge_all}
\end{figure*}

\begin{insightbox}
\textbf{Finding 3:} FRE and FKGL remained stable ($\approx$30 and 14–15), indicating no post-ChatGPT shifts in readability. By contrast, lexical density rose from \textbf{0.55→0.60} and difficult-word ratios increased by \textbf{5–8\%}, implying a gradual enrichment of informational content and technical vocabulary without altering overall text accessibility.
\end{insightbox}

\subsection{(RQ4) AI-Related Research Topics}
\label{sec:RQ4}

\begin{figure}[!t]
  \centering
  \includegraphics[width=\linewidth]{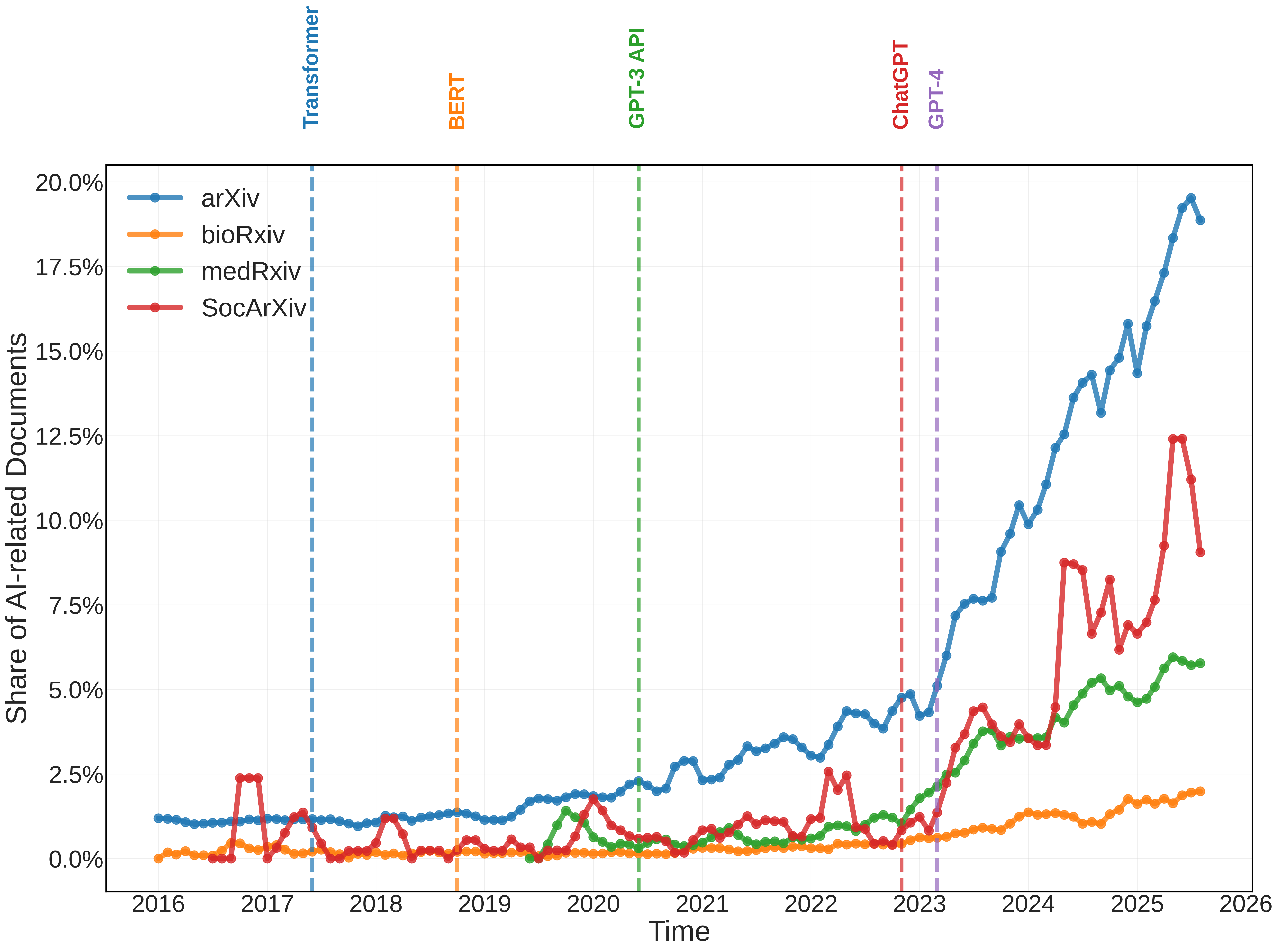}
  \caption{Share of AI-related documents across repositories} %这个可以增大字体 边框变黑 和前面那几个一样
  \label{fig:rq5_aishare}
\end{figure}

The analysis of AI-related topical emergence reveals pronounced shifts in the composition of preprint submissions. \textcolor{teal}{Fig.\ref{fig:rq5_aishare}} shows that the share of AI-flagged documents exhibits a sharp post-2022 inflection, particularly on \textit{arXiv}, where the proportion rises from under 5\% prior to ChatGPT’s release to nearly 20\% by mid-2025. A more gradual but statistically significant increase is observed for \textit{medRxiv} and \textit{SocArXiv}, both of which record sustained upward trajectories in the range of 5–10\%. By contrast, \textit{bioRxiv} displays a comparatively muted trend, with adoption stabilizing around 2\%, suggesting a slower penetration of AI discourse into life-science preprints. Independent-sample \textit{t}-tests confirm that mean adoption rates are significantly higher in the post-ChatGPT period across all platforms, while OLS slope estimates indicate that acceleration is most pronounced for arXiv and SocArXiv, reflecting disciplinary differences in receptivity to AI integration.

The topical decomposition via NMF corroborates these trends, highlighting the diffusion of AI-related themes across distinct domains. The word cloud for arXiv (\textcolor{teal}{Fig.\ref{fig:arxiv_wc}}) is dominated by terms such as “quantum,” “causal,” “learning,” and “functional,” indicating the blending of AI with core areas of physics and computational mathematics. For bioRxiv (\textcolor{teal}{Fig.\ref{fig:biorxiv_wc}}), dominant terms include “cell,” “genes,” “pathways,” and “signaling,” revealing the incorporation of AI methods into genomics and molecular biology. MedRxiv’s cloud (\textcolor{teal}{Fig.\ref{fig:medrxiv_wc}}) is characterized by “covid,” “patients,” “clinical,” and “treatment,” underscoring the role of AI in medical informatics and pandemic-related studies. SocArXiv (\textcolor{teal}{Fig.\ref{fig:socarxiv_wc}}) emphasizes “social,” “language,” “policy,” and “development,” pointing to the uptake of AI tools in the social sciences for text analysis and policy modeling.

\begin{insightbox}
\textbf{Finding 4:} The share of AI-related papers surged after ChatGPT’s release, from $<2\%$ in 2022 to \textbf{$\approx$9\%} in 2025, with strong gains in computer science, bioinformatics, and medical diagnostics. Trend slopes (\(\beta_{\text{post}}\!\approx\!\textbf{0.2}\), \(p<0.01\)) indicate sustained acceleration, confirming that AI topics are diffusing beyond computational fields into life and social sciences.
\end{insightbox}

\subsection{(RQ5) Impacts of ChatGPT Vary Across Disciplines}
\label{sec:RQ5}

\smallskip
\noindent\textbf{a. Disciplinary stratification.} 
The disciplinary stratification reveals that the adoption of AI-related research is highly uneven across fields, both in terms of onset and intensity. In arXiv (\textcolor{teal}{Fig.\ref{fig:arxiv_ridge}}), adoption patterns are highly skewed toward computationally intensive domains. Computer science and quantitative biology display sharp increases in AI-related shares after 2022, with computer science surpassing 50\% of submissions in certain subfields. By contrast, physics and mathematics follow more modest trajectories, remaining below 20\%. This divergence underscores that AI uptake is concentrated in fields already aligned with computational infrastructures and algorithmic practices. In bioRxiv ((\textcolor{teal}{Fig.\ref{fig:biorxiv_ridge}})), the diffusion of AI is broader but similarly uneven. Genomics, neuroscience, and bioinformatics exhibit rapid acceleration post-ChatGPT, highlighting the integration of machine learning into data-rich areas of the life sciences. More traditional domains such as ecology and evolutionary biology show only incremental change, suggesting that methodological dependence on empirical observation slows AI assimilation.

In contrast, medRxiv and SocArXiv reveal more constrained adoption. For medRxiv (\textcolor{teal}{Fig.\ref{fig:medrxiv_ridge}}), growth is concentrated in methodological categories such as health informatics and diagnostics, while clinical and epidemiological research remain largely unaffected, with most disciplines plateauing below 10–15\%. SocArXiv (\textcolor{teal}{Fig.\ref{fig:socarxiv_ridge}}) exhibits the most heterogeneous patterns, where computational social science and political communication show growth, yet traditional sociology and humanities-adjacent fields persist near zero adoption.

\begin{figure}[!t]
  \centering
  \subfigure[Distribution of AI-share deltas and post-ChatGPT slopes]{%
    \includegraphics[width=\linewidth]{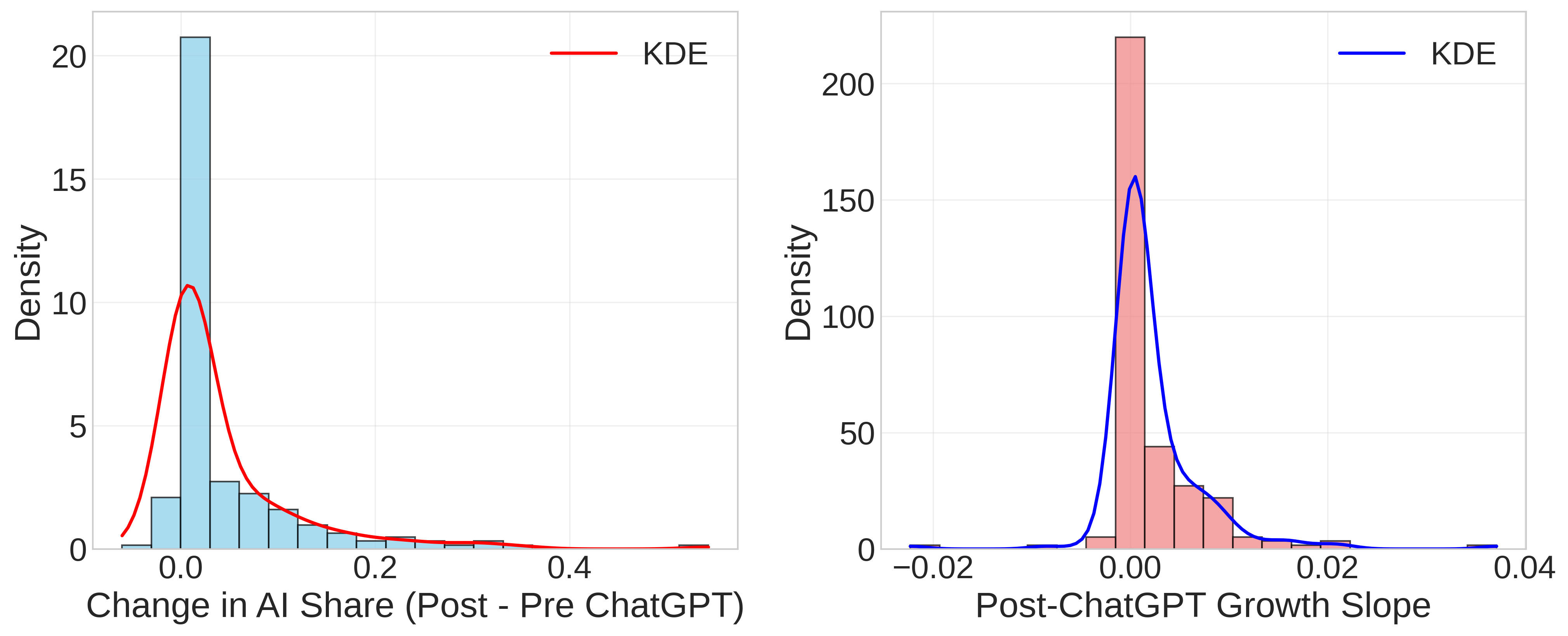}%
    \label{fig:combined_density}}
  \vskip\baselineskip
  \subfigure[Field-level AI adoption trends]{% 同理 这个也黑框
    \includegraphics[width=\linewidth]{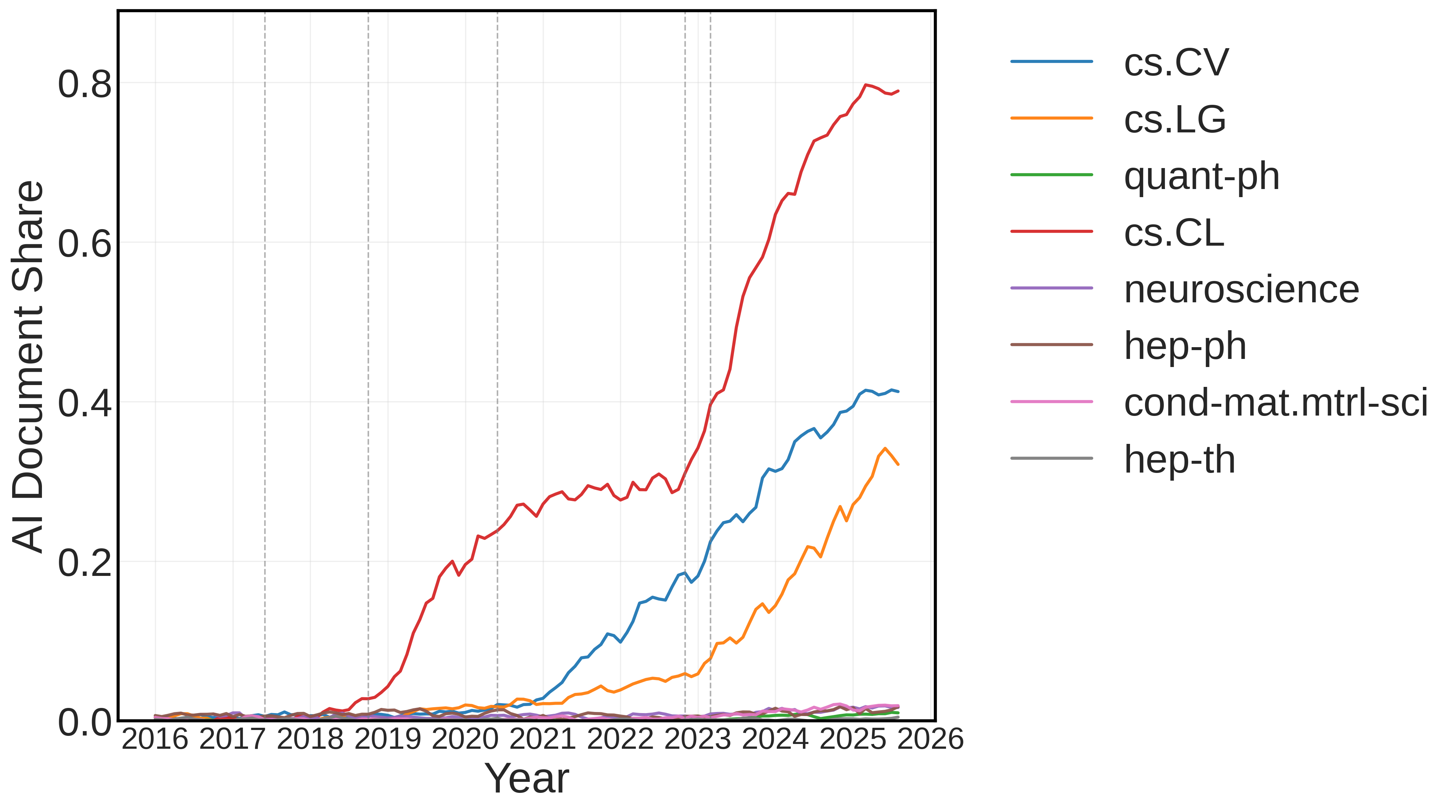}%
    \label{fig:combined_trends}}
  \caption{Cross-disciplinary adoption of AI-related topics. (a) Density plots show the distribution of AI-share changes before and after ChatGPT (left) and the slopes of post-ChatGPT adoption trajectories (right). (b) Field-specific adoption curves illustrate heterogeneous uptake across domains.}
  \label{fig:rq7_combined}
\end{figure}

\smallskip
\noindent\textbf{b. Cross-disciplinary adoption.} 
The cross-disciplinary distributional patterns are summarized in \textcolor{teal}{Fig.\ref{fig:combined_density}}. The left panel illustrates the distribution of changes in AI-related document share before and after the release of ChatGPT, showing that most disciplines exhibit modest deltas centered near zero, while a small number of fields display pronounced positive shifts. This indicates that ChatGPT did not uniformly transform all domains, but rather amplified AI adoption in selected areas. The right panel presents the distribution of post-ChatGPT growth slopes, where the mass is concentrated around zero yet accompanied by a long positive tail. Such a pattern suggests that while the majority of disciplines experienced relatively flat trajectories, a subset of fields underwent accelerated growth after ChatGPT’s introduction.

\smallskip
\noindent\textbf{c. Field-specific adoption.} 
\textcolor{teal}{Fig.\ref{fig:combined_trends}} provides a temporal decomposition of these dynamics across disciplines (top eight). Subfields within computer science, particularly \textsf{cs.CL} and \textsf{cs.CV}, surpassed the 1\% adoption threshold as early as 2019 and subsequently entered an exponential growth regime, reaching 40--80\% of submissions by 2025. In contrast, traditional domains such as high-energy physics (\textsf{hep-ph}) and microbiology remain at single-digit levels, even in the post-ChatGPT period. These heterogeneous trajectories indicate that ChatGPT reinforced pre-existing differences: fields already structurally aligned with AI integration accelerated disproportionately, whereas others showed limited responsiveness.

\begin{insightbox}
\textbf{Finding 5:} AI adoption is highly uneven across disciplines: exceeding \textbf{50\%} in computer science but remaining below \textbf{20\%} in physics and mathematics. Life sciences show intermediate diffusion ($\approx$10\% in genomics and neuroscience), while clinical and social domains lag (<5\%). These contrasts reveal that LLM integration is mediated by disciplinary data cultures and methodological infrastructures rather than uniform technological uptake.
\end{insightbox}

%============================================
\section{Discussions}
\label{sec:discussion}
%============================================
This section integrates empirical results into a unified explanatory framework, considers the limits of causal interpretation, and reflects on the practical consequences for the future of academic publishing.

\subsection{A Coherent Narrative of Change}

The evidence across research questions does not describe isolated phenomena but a connected transformation in the academic ecosystem. When read together, the patterns in submission dynamics, collaboration, linguistic style, topical composition, and disciplinary distribution reveal how LLMs have reconfigured the production cycle of knowledge.

\smallskip
\noindent\textbf{From acceleration to reorganization.}
The patterns observed in RQ1 and RQ2 indicate that the role of LLMs in scholarly publishing extends beyond mere acceleration to a deeper reorganization of academic workflows. The consistent shortening of revision intervals and rising submission activity suggest that generative tools have reduced the cognitive frictions traditionally associated with writing and revision. As these efficiencies accumulate, they reshape how research teams coordinate and allocate intellectual effort: drafting become distributed across contributors and iterative cycles of dissemination occur with increasing frequency. Yet this transformation is uneven across disciplines. Fields that already operate within rapid, data-driven publication cultures (such as the physical and computational sciences) adapt quickly, while more interpretive domains exhibit slower and more selective uptake. These asymmetries imply that LLMs act not as uniform accelerators but as organizational catalysts, amplifying preexisting infrastructures of speed and collaboration.

\smallskip
\noindent\textbf{From reorganization to redirection.} 
While structural reorganization alters how research is produced, its downstream effect is a redirection of what research is pursued. Evidence from RQ4 and RQ5 shows that the post-ChatGPT period has been marked by a rapid expansion of AI-related topics across repositories, indicating that LLMs have influenced not only the mechanics of writing but also the intellectual orientation of scholarship itself. As generative tools became integrated into academic workflows, they also became objects of inquiry, catalyzing new intersections between computational methods and traditional disciplines. This dual role (i.e., instrument and subject) has created a feedback loop in which the availability of LLMs stimulates demand for AI expertise, which in turn encourages further methodological hybridization. Yet this redirection remains asymmetrical. Domains with established quantitative or data-intensive traditions rapidly absorb AI themes into their conceptual mainstream, while fields rooted in qualitative interpretation or normative inquiry incorporate them only at the margins.

\smallskip
\noindent\textbf{From redirection to recomposition.}  
The linguistic shifts documented in RQ3 add another layer to this evolving landscape. On the surface, rising lexical density and higher difficult-word ratios might imply that scientific prose is becoming more complex. Yet, when viewed in conjunction with the thematic redirection identified in RQ4, these patterns reveal a compositional rather than stylistic transformation. What appears as a uniform drift toward “AI-like” writing is, in fact, a reweighting of the scientific corpus itself: more manuscripts now originate from quantitatively oriented communities with inherently dense vocabularies. Thus, the observed change is not merely in how researchers write but in who participates in writing and what kinds of knowledge dominate the publication. This recomposition underscores that LLMs are reshaping the traditional cognitive: altering the linguistic texture of scholarship through shifts in disciplinary representation rather than universal stylistic convergence.

\subsection{Boundaries of Interpretation}
While the cross-RQ suggests plausible mechanisms, several boundaries constrain the strength of our inferences. These boundaries clarify what our analyses can/cannot establish.

\smallskip
\noindent\textbf{Preprints vs. peer-reviewed publications.}
Our dataset is drawn from preprint repositories. Submission acceleration (RQ1–RQ2) therefore reflects author-side behavior, not acceptance. Journals may enforce disclosure requirements or tighten quality standards that alter downstream publication outcomes. Consequently, our findings speak to supply-side activity rather than the full life cycle of scholarly communication.

\smallskip
\noindent\textbf{Abstract-based analysis.}
Readability and linguistic complexity (RQ3) were measured from titles and abstracts. Although this ensures comparability across repositories, abstracts are short and stylized by community norms. What appears as increased difficulty may partly reflect abstract-length constraints or disciplinary conventions. Without full-text validation, we cannot claim comprehensive stylistic transformation.

\smallskip
\noindent\textbf{Topic vs. tool use.}
The rise of AI-related documents (RQ4, RQ5) captures thematic composition, not direct evidence of LLM-assisted writing. Many authors may employ LLMs without explicit mention in titles or abstracts, and conversely, AI-related topics may be studied without LLM support. This limits our ability to disentangle changes in research agendas from changes in writing practices.

\smallskip
\noindent\textbf{Event timing and confounds.}
Interrupted time-series models (RQ1) rely on GPT-3 (2020) and ChatGPT (2022) as intervention points. These coincide with other systemic shocks, including the COVID-19 pandemic and repository policy updates. Although seasonal controls and HAC-robust errors mitigate bias, collinearity and overlapping interventions make strong causal attribution difficult. Our results are consistent with, but do not prove, a direct LLM effect.

\begin{insightbox}
\faLightbulbO~\textbf{Core Insight.} The evidence indicates that LLMs have not transformed academic publishing in a uniform way. Instead, they act as a \textbf{\textit{selective accelerator}}: research cycles are shortened and growth is amplified most strongly in fields already equipped with computational practices. At the same time, shifts in topics and vocabulary can create the appearance of stylistic change. While these patterns are consistent with the influence of generative AI, strong causal claims require caution.
\end{insightbox}

\subsection{Causal Reflection}
The convergent patterns observed across RQs suggest that the release of LLMs has played a role in reshaping scholarly production. However, translating these associations into causal claims requires careful reflection. Our evidence supports certain causal inferences, while leaving others beyond the reach of the current design.

\smallskip
\noindent\textbf{Where causal support is stronger.}
Several findings provide evidence that is consistent with a causal effect of LLM availability. First, the shortening of revision intervals (RQ2) occurs across multiple repositories and coincides temporally with LLM releases. This pattern is difficult to explain by unrelated exogenous shocks, pointing to reduced writing and editing frictions as a plausible causal mechanism. Second, the acceleration in submission volumes following November 2022 (RQ1), particularly in computationally intensive fields, aligns with the supply-side acceleration channel: as LLMs lower the marginal costs of producing manuscripts, aggregate throughput increases. Finally, the fact that changes are selective (RQ5), with computationally oriented fields benefiting the most, suggests that LLMs function more as a catalyst in areas with existing strengths than as a universal driver across all domains.

\smallskip
\noindent\textbf{Where causal claims are weaker.}
Other findings should not be interpreted as direct causal consequences of LLMs. For instance, the rise of AI-related research topics (RQ4) reflects both supply-side and demand-side pressures: the diffusion of LLMs into public discourse, shifts in funding priorities, and policy initiatives all contribute to topic reweighting. Similarly, linguistic changes in abstracts (RQ3) could be by-products of compositional shifts rather than tool use itself. More AI-related submissions naturally raise lexical density and difficult-word ratios, even if authors did not employ LLMs in writing. Moreover, our interrupted time-series models (RQ1) coincide with other systemic events, such as the COVID-19 pandemic, which produced shocks in submission patterns independent of LLM adoption. These confounds caution against attributing all inflection points solely to LLM releases.

\smallskip
\noindent\textbf{Analytical strategies for stronger inference.}
To move beyond simple correlations, we need designs that can approximate what would have happened without LLMs. One option is to use synthetic controls, where the trajectory of a high-exposure field (such as computational linguistics) is compared against a weighted blend of low-exposure fields that together represent its counterfactual. Another option is a staggered difference-in-differences approach~\cite{baker2022much}, which leverages the fact that fields adopted LLM-related topics at different times; this variation allows us to test whether adoption coincides with measurable changes. In addition, placebo interventions (e.g., pretending LLMs appeared in 2018) and robustness checks for repository policy changes help ensure that observed effects are not driven by confounding shocks.

\subsection{Practical Implications}
\label{sec:implications}

Our findings carry deep implications for the structure of knowledge production and scholarly communication.

\smallskip
\noindent\textbf{\ding{172} Shifting boundaries of authorship and intellectual contribution.}
The evidence from collaboration dynamics (RQ2) and linguistic changes (RQ3) indicates that LLMs are blurring the traditional boundaries of authorship. Historically, authorship has implied both the intellectual conception of ideas and the manual production of text. In the LLM era, these two dimensions may increasingly decouple. Drafting and revision, once labor-intensive tasks that signaled ownership of intellectual effort, can now be partially delegated to generative tools. As a result, human contribution may shift toward higher-level orchestration: defining research questions, curating data, interpreting results, and coordinating teams.

This redistribution of effort has several implications. First, collaboration patterns may evolve: if text generation and summarization become less costly, larger teams may form around specialized roles such as prompt engineering, validation, or tool integration, rather than conventional co-writing. Second, the meaning of productivity metrics is destabilized. A rapid increase in per-author output (RQ2) could reflect genuine intellectual acceleration, but it might equally represent efficiency gains from LLM-assisted drafting. Third, the stylistic consistency enabled by LLMs challenges traditional markers of academic identity (e.g., style, tone, and rhetorical voice), raising questions about whether individuality in writing remains a legitimate signal of authorship.

At a normative level, these shifts complicate existing frameworks of credit and accountability. If a manuscript is partly authored by an LLM, should credit be allocated differently among human contributors? If stylistic convergence reduces the visibility of individual authorial voice, how should peer reviewers and readers assess originality? And if productivity becomes mediated by access to advanced tools, does this create inequities in career progression? These questions underline that authorship in the LLM era is no longer simply about “who wrote the text,” but about who contributed intellectual agency in a tool-mediated environment.

\smallskip
\noindent\textbf{\ding{173} Widening inequalities across disciplines.}
Findings from RQ5 show that LLM adoption has not been evenly distributed across fields. Computationally intensive domains, such as computer science, quantitative biology, and bioinformatics, display steep post-ChatGPT growth, while observational or qualitatively oriented disciplines remain largely stagnant. This selective acceleration widens disciplinary divides: some communities gain disproportionate visibility and productivity benefits, while others fall further behind.

These disparities have both structural and institutional roots. Fields with established computational infrastructures and preprint cultures are naturally positioned to integrate LLMs into their workflows, compounding their advantages in speed and output. In contrast, disciplines that rely on slower, resource-intensive methods, such as clinical trials or archival research, lack similar opportunities to benefit from generative tools. The result is a  gap not only in submission volume but also in how quickly knowledge can be disseminated.

At a global level, these divides risk reinforcing geographic and institutional inequalities. Researchers in low-resource settings, already disadvantaged by limited access to computational infrastructure, may now face a “double burden”: slower methodological cycles in their home disciplines combined with fewer opportunities to leverage LLM acceleration. Over time, such dynamics could concentrate scholarly influence in high-adoption domains and well-resourced institutions, reshaping the balance of knowledge production.

\smallskip
\noindent\textbf{\ding{174} Normative challenges for governance.}
The integration of LLMs into research practice raises governance challenges that go beyond technology to the core norms of science. As revision cycles accelerate (RQ2) and AI-assisted writing becomes common (RQ3), distinguishing human from machine-generated text is increasingly difficult. Without clear disclosure rules, repositories and journals risk eroding trust in the scholarly record. Governance bodies should therefore adopt standards similar to conflict-of-interest statements, requiring authors to state whether and how LLMs were used.

As RQ2 and RQ3 show, intellectual contribution boundaries are shifting. When LLMs generate parts of a manuscript, responsibility for errors, bias, or plagiarism becomes unclear. Existing authorship frameworks that assume human control are no longer sufficient. Updated guidelines must define legitimate authorship, clarify the role of AI assistance, and specify accountability for the final work.

Uneven LLM adoption across disciplines (RQ5) also raises fairness concerns. Fields with advanced computational infrastructure gain speed advantages, while others may lag. If evaluation systems for tenure, funding, or publication ignore these disparities, structural inequalities may deepen. Governance in the LLM era must therefore sustain trust, clarify responsibility, and ensure fairness across disciplines.

\begin{insightbox}
\faThumbsOUp~\textbf{Implications.} LLMs are reshaping the \textbf{\textit{economy of scholarly production}} rather than merely accelerating writing. Authorship is being redefined toward orchestration and oversight, collaboration is expanding along new functional lines, and disciplinary inequalities are widening. The challenge ahead lies not in adoption itself but in ensuring accountability, transparency, and equity.
\end{insightbox}

% \faThumbsOUp

% %============================================
% \section{Related Work}
% \label{sec:rw}
% %============================================

%============================================
\section{Conclusion}
\label{sec:conclusion}
%============================================

This study provides large-scale empirical evidence on how GenAI is reshaping academic publishing. Using over two million preprints across four repositories, we show that LLMs have accelerated submission and revision cycles (RQ1–RQ2), introduced subtle increases in linguistic complexity (RQ3), and amplified the visibility of AI-related topics (RQ4), while disproportionately benefiting computationally intensive fields (RQ5). Rather than acting as a universal driver of change, LLMs function as selective catalysts that reinforce existing strengths and widen disciplinary divides. These dynamics highlight both opportunities and risks: while generative AI can enhance efficiency and expand thematic horizons, it also raises challenges for authorship, fairness, and governance. Our findings lay an empirical foundation for tracking these transformations and call for careful design of policies and metrics to ensure that the evolving landscape of AI-enabled scholarship remains credible and equitable.

% %============================================
% \section{Ethical considerations}
% \label{sec:ethical}
% %============================================

% All data used in this study were collected from openly accessible preprint repositories (\textit{arXiv}, \textit{bioRxiv}, \textit{medRxiv}, and \textit{SocArXiv}) via their official APIs, in full compliance with the platforms' terms of service. No personally identifiable information beyond publicly listed author names was retrieved, and all analyses were conducted at the aggregate level to prevent re-identification of individual scholars. The study does not seek to evaluate or judge specific authors, institutions, or manuscripts, but rather to identify broad patterns in scholarly publishing associated with the diffusion of generative AI. Potential biases stemming from incomplete metadata, heterogeneous platform coverage, or uneven disclosure practices are acknowledged and discussed as methodological limitations. By focusing on publicly available and de-identified records, we ensure that the study adheres to established norms of research integrity and minimizes risks of privacy infringement or reputational harm.

%============================================
\bibliographystyle{unsrt}
\bibliography{bib/bib}
%============================================

%-----------------------------------------
%\section*{Acknowledgement}
%-----------------------------------------

%This work does not raise any ethical issues. All the data sources used are publicly accessible websites and APIs. 

%\smallskip

%\noindent\textbf{Data Use Disclaimer.} This work does not raise any ethical issues. All the data we crawl from Snapshot are open-released and free to use with CC0 licenses. We strive to maintain the accuracy of all data that we crawl from Snapshot and declare that the data will not be used for any commercial purposes. 

\appendix

\noindent\textit{Table~\ref{tab-llm-summary}} provides an overview of the rapidly evolving ecosystem that underpins AI-augmented scientific research. It maps the interconnections among four key dimensions: LLMs, toolchains, evaluation benchmarks, and open data infrastructures, collectively defining the technical foundation of the GenAI research landscape.

\begin{table*}[!hbt]
\centering
\caption{Ecosystem of LLMs, Tools, Benchmarks, and Data Infrastructures for Scientific Research}
\label{tab-llm-summary}
\begin{threeparttable}
\renewcommand{\arraystretch}{1.4}
\resizebox{0.99\linewidth}{!}{
\begin{tabular}{ll|ccc|*{5}{>{\centering\arraybackslash}p{0.7cm}}|*{6}{>{\centering\arraybackslash}p{0.7cm}}|*{4}{>{\centering\arraybackslash}p{0.7cm}}}

\toprule
& \multicolumn{1}{c}{ \multirow{2}{*}{\textit{\textbf{LLM / System}}} } 
& \multicolumn{3}{c}{\textit{\textbf{Operational Features}}} 
& \multicolumn{5}{c}{\textit{\textbf{Functionalities (Research-Oriented)}}} 
& \multicolumn{6}{c}{\textit{\textbf{Non-Functionalities}}} 
& \multicolumn{4}{c}{\textbf{\textit{Performance Snapshot}}} \\
\cmidrule{3-19}
& & \textit{\textbf{Access}} & \textit{\textbf{License}} & \textit{\textbf{Modality}} 
& \rotatebox{90}{\textit{\textbf{Search}}} 
& \rotatebox{90}{\textit{\textbf{Summarize}}} 
& \rotatebox{90}{\textit{\textbf{Cite-aware}}} 
& \rotatebox{90}{\textit{\textbf{Code-asst}}} 
& \rotatebox{90}{\textit{\textbf{Data-extract}}} 
& \rotatebox{90}{\textit{\textbf{Reproducible}}} 
& \rotatebox{90}{\textit{\textbf{Transparent}}} 
& \rotatebox{90}{\textit{\textbf{Auditable}}} 
& \rotatebox{90}{\textit{\textbf{Safety filters}}} 
& \rotatebox{90}{\textit{\textbf{Offline}}} 
& \rotatebox{90}{\textit{\textbf{Local fine-tune}}} 
& \rotatebox{90}{\textit{\textbf{Context}}} 
& \rotatebox{90}{\textit{\textbf{Cost}}} 
& \rotatebox{90}{\textit{\textbf{Speed}}} 
& \rotatebox{90}{\textit{\textbf{Multilingual}}} \\
\midrule

\multirow{16}{*}{\rotatebox{90}{\textit{\textbf{LLMs / Systems}}}} 
& GPT-class (hosted) & API/Web & Closed & Text+Vision & \cmark & \cmark & \cmark & \cmark & \cmark & \xmark & \xmark & \xmark & \cmark & \xmark & \xmark & High & Mid & High & \cmark \\
& Claude-class (hosted) & API/Web & Closed & Text+Vision & \cmark & \cmark & \cmark & \cmark & \cmark & \xmark & \xmark & \xmark & \cmark & \xmark & \xmark & High & Mid & High & \cmark \\
& Gemini-class (hosted) & API/Web & Closed & Multimodal & \cmark & \cmark & \cmark & \cmark & \cmark & \xmark & \xmark & \xmark & \cmark & \xmark & \xmark & High & Mid & High & \cmark \\
& Copilot-style (IDE) & API/IDE & Closed & Text+Code & \xmark & \cmark & \xmark & \cmark & \cmark & \xmark & \xmark & \xmark & \cmark & \xmark & \xmark & Med & Mid & High & \cmark \\
& Perplexity-style (RAG) & Web/API & Closed & Text+Web & \cmark & \cmark & \cmark & \xmark & \cmark & \xmark & \xmark & \xmark & \cmark & \xmark & \xmark & Med & Mid & High & \cmark \\
& Elicit-style (literature) & Web/API & Closed & Text & \cmark & \cmark & \cmark & \xmark & \cmark & \xmark & \xmark & \xmark & \cmark & \xmark & \xmark & Med & Mid & Med & \cmark \\
& scite-style (citation) & Web/API & Closed & Text & \cmark & \cmark & \cmark & \xmark & \cmark & \xmark & \xmark & \xmark & \cmark & \xmark & \xmark & Med & Mid & Med & \cmark \\
& Semantic-Scholar AI & Web/API & Closed & Text & \cmark & \cmark & \cmark & \xmark & \cmark & \xmark & \xmark & \xmark & \cmark & \xmark & \xmark & Med & Low & Med & \cmark \\
& Llama-class (local) & API/Local & Open & Text(+Vision) & \cmark & \cmark & \xmark & \cmark & \cmark & \cmark & \cmark & \cmark & \xmark & \cmark & \cmark & Med & Low & Med & \cmark \\
& Mistral-class (local) & API/Local & Open & Text & \cmark & \cmark & \xmark & \cmark & \cmark & \cmark & \cmark & \cmark & \xmark & \cmark & \cmark & Med & Low & Med & \cmark \\
& T5/Flan (fine-tune) & Local & Open & Text & \xmark & \cmark & \xmark & \cmark & \cmark & \cmark & \cmark & \cmark & \xmark & \cmark & \cmark & Low & Low & Med & \xmark \\
& SciBERT/Longformer & Local & Open & Text (long) & \xmark & \cmark & \xmark & \xmark & \cmark & \cmark & \cmark & \cmark & \xmark & \cmark & \cmark & High & Low & Med & \xmark \\
& RetNet/RWKV (eff.) & Local & Open & Text & \xmark & \cmark & \xmark & \xmark & \cmark & \cmark & \cmark & \cmark & \xmark & \cmark & \cmark & High & Low & High & \xmark \\
& Agentic pipelines & API/Local & Mixed & Multi-step & \cmark & \cmark & \cmark & \cmark & \cmark & \xmark & \xmark & \xmark & \cmark & \xmark & \cmark & Med & Var & Var & \cmark \\
& RAG frameworks & API/Local & Open & Text+Docs & \cmark & \cmark & \cmark & \xmark & \cmark & \cmark & \cmark & \cmark & \xmark & \cmark & \cmark & High & Low & High & \cmark \\
& Manubot-like (writing) & Local/Web & Open & Text/PDF & \xmark & \cmark & \cmark & \xmark & \cmark & \cmark & \cmark & \cmark & \xmark & \cmark & \cmark & Med & Low & Med & \xmark \\
\midrule

& \multicolumn{1}{c}{\rotatebox{0}{\textit{\textbf{Projects/Tools}}}} 
&  \multicolumn{2}{c}{\rotatebox{0}{\textit{\textbf{Focus}}}}  & \multicolumn{1}{c|}{\rotatebox{0}{\textit{\textbf{Note}}}}   
& \multicolumn{1}{c}{} & 
\multicolumn{4}{c|}{\rotatebox{0}{\textit{\textbf{Projects/Feeds}}}} & \multicolumn{3}{c}{\rotatebox{0}{\textit{\textbf{Coverage}}}} & \multicolumn{7}{c}{\rotatebox{0}{\textit{\textbf{Note / URL}}}}\\
\midrule

\multirow{12}{*}{ \rotatebox{90}{\textit{\textbf{Tools \textcolor{gray}{\&} Launchpad}}}} 
& Overleaf & \multicolumn{2}{c}{LaTeX authoring} & \multicolumn{1}{c|}{Collab writing}  & 
& \multicolumn{4}{c|}{GitHub} & \multicolumn{3}{c}{Code, Issues, Releases} & \multicolumn{7}{c}{\url{https://github.com}} \\
& Manubot & \multicolumn{2}{c}{Executable papers} & \multicolumn{1}{c|}{CI builds}  & 
& \multicolumn{4}{c|}{OpenAlex API} & \multicolumn{3}{c}{Bibliometrics} & \multicolumn{7}{c}{\url{https://openalex.org}} \\
& Zotero & \multicolumn{2}{c}{Reference mgmt} & \multicolumn{1}{c|}{Citation graph}  & 
& \multicolumn{4}{c|}{Crossref API} & \multicolumn{3}{c}{DOIs, metadata} & \multicolumn{7}{c}{\url{https://api.crossref.org}} \\
& scite.ai & \multicolumn{2}{c}{Citation intent} & \multicolumn{1}{c|}{Support/contrast}  & 
& \multicolumn{4}{c|}{PubMed / E-utilities} & \multicolumn{3}{c}{Biomed refs} & \multicolumn{7}{c}{\url{https://pubmed.ncbi.nlm.nih.gov}} \\
& Hypothesis & \multicolumn{2}{c}{Web annotation} & \multicolumn{1}{c|}{Open notes}  & 
& \multicolumn{4}{c|}{arXiv API} & \multicolumn{3}{c}{Preprints} & \multicolumn{7}{c}{\url{https://export.arxiv.org/api}} \\
& Ray / vLLM & \multicolumn{2}{c}{Serving infra} & \multicolumn{1}{c|}{Scale inference}  & 
& \multicolumn{4}{c|}{bioRxiv/medRxiv API} & \multicolumn{3}{c}{Bio/Med preprints} & \multicolumn{7}{c}{\url{https://api.biorxiv.org}} \\
& DVC / Weights\&Biases & \multicolumn{2}{c}{Exp tracking} & \multicolumn{1}{c|}{Reproducibility}  & 
& \multicolumn{4}{c|}{SocArXiv / OSF API} & \multicolumn{3}{c}{SocSci preprints} & \multicolumn{7}{c}{\url{https://api.osf.io/v2}} \\
& FastAPI / LangChain & \multicolumn{2}{c}{RAG pipeline} & \multicolumn{1}{c|}{Prod glue}  & 
& \multicolumn{4}{c|}{Dimensions / Lens} & \multicolumn{3}{c}{Citations/affils} & \multicolumn{7}{c}{\url{https://www.dimensions.ai}} \\
& Manifold / OpenReview & \multicolumn{2}{c}{Peer review stack} & \multicolumn{1}{c|}{Community review}  & 
& \multicolumn{4}{c|}{CORE / BASE} & \multicolumn{3}{c}{Repositories} & \multicolumn{7}{c}{\url{https://core.ac.uk}} \\
& Prodigy / Label Studio & \multicolumn{2}{c}{Annotation} & \multicolumn{1}{c|}{Data curation}  & 
& \multicolumn{4}{c|}{OpenAIRE} & \multicolumn{3}{c}{Scholarly graph} & \multicolumn{7}{c}{\url{https://explore.openaire.eu/}} \\
& JATS/Docutils toolchain & \multicolumn{2}{c}{XML/PDF ops} & \multicolumn{1}{c|}{Parse/convert}  & 
& \multicolumn{4}{c|}{Unpaywall} & \multicolumn{3}{c}{Open access} & \multicolumn{7}{c}{\url{https://unpaywall.org}} \\
& COPE / ICMJE templates & \multicolumn{2}{c}{Governance} & \multicolumn{1}{c|}{Disclosure norms}  & 
& \multicolumn{4}{c|}{RetractionWatch} & \multicolumn{3}{c}{Corrections} & \multicolumn{7}{c}{\url{https://retractionwatch.com}} \\
\midrule

& \multicolumn{1}{c}{\rotatebox{0}{\textit{\textbf{Benchmarks/Datasets}}}} 
&  \multicolumn{2}{c}{\rotatebox{0}{\textit{\textbf{Task focus}}}}  & \multicolumn{1}{c|}{\rotatebox{0}{\textit{\textbf{Split/Note}}}}   
& \multicolumn{1}{c}{} & 
\multicolumn{4}{c|}{\rotatebox{0}{\textit{\textbf{Benchmarks/Datasets (More)}}}} & \multicolumn{3}{c}{\rotatebox{0}{\textit{\textbf{Domain}}}} & \multicolumn{6}{c}{\rotatebox{0}{\textit{\textbf{Note / URL}}}}\\
\midrule

\multirow{12}{*}{ \rotatebox{90}{\textit{\textbf{Datasets \textcolor{gray}{\&} Benchmarks}}}} 
& MMLU & \multicolumn{2}{c}{Multidomain QA} & \multicolumn{1}{c|}{57 subjects; public}  & 
& \multicolumn{4}{c|}{LongBench / InfiniteBench} & \multicolumn{3}{c}{Long-context} & \multicolumn{7}{c}{\url{https://github.com/hendrycks/test}} \\
& PubMedQA / BioASQ & \multicolumn{2}{c}{Biomedical QA} & \multicolumn{1}{c|}{Abstract-based; public}  & 
& \multicolumn{4}{c|}{SciEval / SciCoT} & \multicolumn{3}{c}{Scientific writing/QA} & \multicolumn{7}{c}{\url{https://pubmedqa.github.io}} \\
& PaperBench (arXiv-based) & \multicolumn{2}{c}{Paper understanding} & \multicolumn{1}{c|}{Long PDF; public}  & 
& \multicolumn{4}{c|}{ArXivCS / ScholarBench} & \multicolumn{3}{c}{Citations, claims} & \multicolumn{7}{c}{\url{https://github.com/scholarbench}} \\
& GSM8K / MATH & \multicolumn{2}{c}{Math reasoning} & \multicolumn{1}{c|}{Chain-of-thought}  & 
& \multicolumn{4}{c|}{HumanEval / MBPP} & \multicolumn{3}{c}{Code} & \multicolumn{7}{c}{\url{https://github.com/openai/human-eval}} \\
& Qasper / NarrativeQA & \multicolumn{2}{c}{Doc QA / long QA} & \multicolumn{1}{c|}{Evidence-grounded}  & 
& \multicolumn{4}{c|}{SciRepEval} & \multicolumn{3}{c}{Reproducibility} & \multicolumn{7}{c}{\url{https://allenai.org/open-datar}} \\
& MS MARCO / BEIR & \multicolumn{2}{c}{Retrieval/RAG} & \multicolumn{1}{c|}{Open-domain}  & 
& \multicolumn{4}{c|}{LongVideoBench (if multimodal)} & \multicolumn{3}{c}{Vision+Text} & \multicolumn{7}{c}{\url{https://github.com/beir-cellar/beir}} \\
& MedMCQA / MedQA-USMLE & \multicolumn{2}{c}{Clinical QA} & \multicolumn{1}{c|}{Exam-style; public}  & 
& \multicolumn{4}{c|}{BioNLI / ChemProt} & \multicolumn{3}{c}{Bio IE/NLI} & \multicolumn{7}{c}{\url{https://medmcqa.github.io}} \\
& DocLayNet / PubTables-1M & \multicolumn{2}{c}{Doc \& table parsing} & \multicolumn{1}{c|}{Layout+tables}  & 
& \multicolumn{4}{c|}{ChartQA / PlotQA} & \multicolumn{3}{c}{Chart reasoning} & \multicolumn{7}{c}{\url{https://www.microsoft.com/en-us/research/publication/pubtables-1m/}} \\
& GovReport / ArXivSumm & \multicolumn{2}{c}{Long summarization} & \multicolumn{1}{c|}{Report/papers}  & 
& \multicolumn{4}{c|}{S2ORC (subset tasks)} & \multicolumn{3}{c}{Scholarly corpora} & \multicolumn{7}{c}{\url{https://gov-report-data.github.io}} \\
& HotpotQA / NQ-Open & \multicolumn{2}{c}{Multi-hop QA} & \multicolumn{1}{c|}{Open-domain}  & 
& \multicolumn{4}{c|}{TRECCAR / TREC-COVID} & \multicolumn{3}{c}{IR / domain search} & \multicolumn{7}{c}{\url{https://hotpotqa.github.io}} \\
& DocVQA / InfographicVQA & \multicolumn{2}{c}{Doc VQA} & \multicolumn{1}{c|}{Text+layout}  & 
& \multicolumn{4}{c|}{TextCaps / TextVQA} & \multicolumn{3}{c}{Vision+text} & \multicolumn{7}{c}{\url{https://docvqa.org}} \\
& Evidence Inference & \multicolumn{2}{c}{Claim-evidence} & \multicolumn{1}{c|}{Clinical trials}  & 
& \multicolumn{4}{c|}{SciERC} & \multicolumn{3}{c}{Entity/relation} & \multicolumn{7}{c}{\url{https://evidence-inference.ebm-nlp.com}} \\
\midrule

& \multicolumn{1}{c}{\rotatebox{0}{\textit{\textbf{Platforms/Infra}}}} 
&  \multicolumn{2}{c}{\rotatebox{0}{\textit{\textbf{Function}}}}  & \multicolumn{1}{c|}{\rotatebox{0}{\textit{\textbf{Scope}}}}   
& \multicolumn{1}{c}{} & 
\multicolumn{4}{c|}{\rotatebox{0}{\textit{\textbf{Platforms/Infra (More)}}}} & \multicolumn{3}{c}{\rotatebox{0}{\textit{\textbf{Coverage}}}} & \multicolumn{6}{c}{\rotatebox{0}{\textit{\textbf{Note / URL}}}}\\
\midrule

\multirow{12}{*}{ \rotatebox{90}{\textit{\textbf{Platforms \textcolor{gray}{\&} Infrastructures}}}} 
& OpenAlex & \multicolumn{2}{c}{Scholarly graph} & \multicolumn{1}{c|}{Works, authors, venues}  & 
& \multicolumn{4}{c|}{Crossref / DataCite} & \multicolumn{3}{c}{DOIs, datasets} & \multicolumn{7}{c}{\url{https://openalex.org}} \\
& Dimensions / Lens & \multicolumn{2}{c}{Discovery/analytics} & \multicolumn{1}{c|}{Grants, policy links}  & 
& \multicolumn{4}{c|}{CORE / BASE / OpenAIRE} & \multicolumn{3}{c}{IR aggregators} & \multicolumn{7}{c}{\url{https://www.base-search.net}} \\
& Unpaywall & \multicolumn{2}{c}{OA resolver} & \multicolumn{1}{c|}{OA flags, legal sources}  & 
& \multicolumn{4}{c|}{Retraction Watch} & \multicolumn{3}{c}{Integrity/ethics} & \multicolumn{7}{c}{\url{https://unpaywall.org}} \\
& arXiv/bioRxiv/medRxiv/SocArXiv  & \multicolumn{2}{c}{Preprint hosts} & \multicolumn{1}{c|}{APIs, metadata}  & 
& \multicolumn{4}{c|}{PubMed / E-utilities} & \multicolumn{3}{c}{Biomed index} & \multicolumn{7}{c}{\url{https://export.arxiv.org/api}} \\
& ORCID / ROR & \multicolumn{2}{c}{Identity/affiliation} & \multicolumn{1}{c|}{Author/Inst. IDs}  & 
& \multicolumn{4}{c|}{OpenAlex Institutions} & \multicolumn{3}{c}{Affiliation graph} & \multicolumn{7}{c}{\url{https://orcid.org}} \\
& Semantic Scholar API & \multicolumn{2}{c}{Paper search/API} & \multicolumn{1}{c|}{Abstracts, citations}  & 
& \multicolumn{4}{c|}{Google Scholar (UI)} & \multicolumn{3}{c}{Discovery} & \multicolumn{7}{c}{\url{https://api.semanticscholar.org}} \\
& OpenReview & \multicolumn{2}{c}{Peer-review platform} & \multicolumn{1}{c|}{Submissions, reviews}  & 
& \multicolumn{4}{c|}{HotCRP (conf mgmt)} & \multicolumn{3}{c}{Workflow} & \multicolumn{7}{c}{\url{https://openreview.net}} \\
& Zenodo / Figshare & \multicolumn{2}{c}{Data/code deposit} & \multicolumn{1}{c|}{DOI assignment}  & 
& \multicolumn{4}{c|}{Dataverse} & \multicolumn{3}{c}{Hosted datasets} & \multicolumn{7}{c}{\url{https://zenodo.org}} \\
& Europe PMC & \multicolumn{2}{c}{Full-text index} & \multicolumn{1}{c|}{OA life sciences}  & 
& \multicolumn{4}{c|}{Sherpa/RoMEO} & \multicolumn{3}{c}{Publisher policies} & \multicolumn{7}{c}{\url{https://europepmc.org}} \\
& DBLP & \multicolumn{2}{c}{CS bibliography} & \multicolumn{1}{c|}{Authors, venues}  & 
& \multicolumn{4}{c|}{HAL/TEL} & \multicolumn{3}{c}{French OA repos} & \multicolumn{7}{c}{\url{https://dblp.org}} \\
& Crossref Event Data & \multicolumn{2}{c}{Citation events} & \multicolumn{1}{c|}{Mentions, links}  & 
& \multicolumn{4}{c|}{Altmetric/PlumX} & \multicolumn{3}{c}{Attention signals} & \multicolumn{7}{c}{\url{https://www.crossref.org/services/event-data}} \\
& Web of Science/Scopus & \multicolumn{2}{c}{Abstracting \& indexing} & \multicolumn{1}{c|}{Closed; analytics}  & 
& \multicolumn{4}{c|}{OpenAlex Snapshots} & \multicolumn{3}{c}{Bulk data} & \multicolumn{7}{c}{\url{https://www.elsevier.com/solutions/scopus}} \\
\bottomrule
\end{tabular}
}
\vspace{0.2em}
\begin{tablenotes}
\footnotesize
\item[] \textbf{Legend:} Access = API/Web/Local; License = Open/Closed/Mixed; Modality = Text/Code/Vision/Multimodal.
\item[] \textbf{Functions:} Search (literature/web), Summarize (papers/sections), Cite-aware (references/DOIs), Code-asst (analysis scripts), Data-extract (tables/figures).
\item[] \textbf{Non-functional:} Reproducible, Transparent (provenance), Auditable (logs), Safety filters (guardrails), Offline (air-gapped), Local fine-tune.
\item[] \textbf{Performance:} Context (relative length), Cost (relative rate), Speed (latency band), Multilingual (baseline coverage). Items are representative.
\end{tablenotes}
\end{threeparttable}
\end{table*}

\end{document}